\documentclass[10pt,draft]{article}

\usepackage{amssymb,amsfonts,amsmath,curves}

\newcommand{\NN}{\mathbb{N}}
\newcommand{\ZZ}{\mathbb{Z}}
\newcommand{\MM}{\mathcal{M}}

\newcommand{\FF}{\mathbb{F}_{2}}

\newcommand{\RR}{\mathbb{R}}

\newcommand{\Id}{\mathrm{Id}}
\newcommand{\Sn}{{S}_{n}}
\newcommand{\SE}{\mathcal{S}_{n}}
\newcommand{\Cn}{\mathcal{C}_{n}}

\newcommand{\qed}{\hbox{}\nobreak\hfill\vrule width 1.4mm height 1.4mm depth 0mm
    \par \goodbreak \smallskip}

\newcommand{\leqs}{\leqslant}
\newcommand{\geqs}{\geqslant}

\newtheorem{theorem}{Theorem}
\newtheorem{proposition}[theorem]{Proposition}
\newtheorem{lemma}[theorem]{Lemma}
\newtheorem{corollary}[theorem]{Corollary}
\newtheorem{example}[theorem]{Example}
\newtheorem{definition}{Definition}

\newenvironment{proof}{\noindent{\it Proof.} \rm }{\hfill\qed}

\begin{document}

\title{A family of metrics on contact structures based on edge 
ideals\thanks{This work has been partially supported by the Spanish 
DGES, grant BFM2000-1113-C02-01.}}

\author{M. Llabr\'es, F. Rossell\'o\\[1ex]
{\small Dept.\ of Mathematics and Computer Science,}\\
{\small Research Institute of Health Science  (IUNICS),}\\
{\small University of the Balearic Islands,}\\
{\small 07122 Palma de Mallorca (Spain)}\\
{\small \emph{E-mail:} \{\texttt{merce.llabres,cesc.rossello}\}\texttt{@uib.es}}}

\maketitle

\begin{abstract}
The measurement of the similarity of RNA secondary structures, and in 
general of contact structures, of a fixed length has several specific 
applications.  For instance, it is used in the analysis of the 
ensemble of suboptimal secondary structures generated by a given 
algorithm on a given RNA sequence, and in the comparison of the 
secondary structures predicted by different algorithms on a given RNA 
molecule.  It is also a useful tool in the quantitative study of 
sequence-structure maps.  A way to measure this similarity is by means 
of metrics.  In this paper we introduce a new class of metrics 
$d_{m}$, $m\geq 3$, on the set of all contact structures of a fixed 
length, based on their representation by means of edge ideals in a 
polynomial ring.  These metrics can be expressed in terms of Hilbert 
functions of monomial ideals, which allows the use of several public 
domain computer algebra systems to compute them.  We study some 
abstract properties of these metrics, and we obtain explicit 
descriptions of them for $m=3,4$ on arbitrary contact structures and 
for $m=5,6$ on RNA secondary structures.  \smallskip

\end{abstract}

\section{Introduction}

As it is well known, in the cell and \textsl{in vitro} RNA molecules 
and proteins fold into three-dimensional structures, which determine 
their biochemical function.  A central problem in 
molecular biology is the study of these structures, their prediction 
and comparison.  As different levels of precision are suitable for 
different problems, we can sometimes forget about the detailed 
description of the three-dimensional structure of a biopolymer and 
simply focus our attention on what has been called its \emph{contact 
structure}: the set of all pairs of monomers (nucleotides in RNA 
molecules, aminoacids in proteins) that are spatial neighbors in the 
three-dimensional structure \cite{Dill0}.  If we assume the monomers 
numbered from 1 to $n$ along the backbone of the polymer, then a 
contact structure can be understood as an undirected graph without 
multiple edges or self-loops with set of nodes $\{1,\ldots,n\}$: its 
edges are consistently called \emph{contacts} and its number $n$ of 
nodes its \emph{length}. 

The secondary structures of RNA molecules form a special class of 
contact structures.  In them, contacts represent the hydrogen bonds 
between pairs of bases that held together the three-dimensional 
structure.  A hydrogen bond can only form between bases that are 
several positions apart in the chain, but we shall not take this 
restriction into account here and we shall  impose that a contact 
can only exist between non-consecutive bases.  A restriction is added 
to the definition of \emph{RNA secondary structure}: a base can only 
pair with at most one base.  This restriction is called the 
\emph{unique bonds condition} and it is specific of  secondary 
structures.  It is usual to impose a further restriction on
 RNA secondary structures, by 
forbidding  the existence of (\emph{pseudo})\emph{knots}: a 
contact between bases at the $i$th and $j$th positions in the backbone 
cannot coexist with a contact between bases at the $k$th and $l$th 
positions  if $i<k<j<l$.  This restriction has its 
origin in the first dynamic programming methods to predict RNA 
secondary structures \cite{WS,Zuk, ZukS}, but since real RNA structures 
can contain knots, which are moreover important structural elements in many RNA 
molecules, and their existence does not compromise our models, we 
shall not impose this restriction  here.

Contact structures with unique bonds can also be used to represent the 
basic building blocks of protein structures, like $\alpha$-helixes, 
$\beta$-sheets and $\beta$ and $\Omega$-turns (called thus 
\emph{protein secondary structures}), which are also held together by 
means of hydrogen bonds between non-consecutive   aminoacids.

But, beyond secondary structures, the representation of the 
neighborhood in three-dimensio\-nal structures of RNA molecules and 
proteins needs contact structures without unique bonds.  The full 
three-dimensional structure of RNA molecules contains contacts that 
violate the unique bonds condition, like base triplets and guanine 
platforms \cite{tert,motif}. And in the tertiary structure of a 
protein, represented for instance by means of a self-avoiding walk in 
a lattice (i.e., a path in $\NN^3$ that does not visit the same node 
more than once \cite{SAW}), one aminoacid can be  spatial neighbor 
of several aminoacids \cite{Dill1,Dill2}.  But even in this general 
case, the existence of contacts between pairs of monomers that are 
next to each other in the backbone is still forbidden in contact 
structures, because their 
spatial closeness can be understood as a consequence of their position in 
the backbone.

As we mentioned, an important problem in molecular biology is the 
comparison of the three-dimensional structures formed by RNA molecules 
and proteins, because it is assumed that a preserved three-dimensional 
structure corresponds to a preserved function.  Moreover, the 
measurement of the similarity of contact structures on biopolymers of 
a fixed length has an interest in itself.  For instance, it can be 
used in the analysis of the ensemble of suboptimal solutions provided 
by a given algorithm, like for instance Zuker's algorithm 
\cite{ZukSc}, to the problem of determining the secondary structure of 
a given RNA molecule; see \cite{Zuk,ZukM}.  It can also be used to 
compare the output of different prediction algorithms applied to the 
same RNA molecule or protein, to assess their performance.  This 
similarity measurement lies also at the basis of the study of the 
mapping that assigns to each RNA molecule or protein the structure it 
folds into \cite{Font93,Schu94} and it can be used in the study of 
phenotype spaces \cite{Font98}.

The similarity of contact structures can be quantified by means of 
metrics  on the set of all contact structures of a given 
length.  For instance, with the purpose of comparing suboptimal 
solutions to the RNA secondary structure prediction problem in order to reduce 
the number of alternate structures obtained by his algorithm, Zuker 
introduced from the very beginning its metric $d_{Z}$ \cite{ZukSc,Zuk} 
and more recently the  \emph{mountain metrics} \cite{ZukM}.  
Tree editing distances have also been used in this context 
\cite{Hof1,ZukM,Shap}.

Reidys and Stadler defined in their seminal paper \cite{RS96} on 
algebraic models of biopolymer structures three metrics on RNA 
secondary structures of fixed length $n$, based on their 
representations as involutions and as permutation subgroups, and on 
Magarshak's matrix representation \cite{Mag}, and they discussed their 
biophysical relevance.  These metrics have been recently analyzed from 
the mathematical point of view  \cite{JJC,Ros03}.

Since their models cannot be used to represent in a one-to-one way 
contact structures without unique bonds, Reidys and Stadler's metrics 
cannot be extended to the set of arbitrary contact structures of a 
fixed length.  In this paper we overcome this drawback, by switching 
from subgroups of the symmetric group $S_{n}$ to monomial ideals of a 
polynomial ring in $n$ variables.  More specifically, we represent a 
contact structure by means of its \emph{edge ideal}.  Edge ideals are 
a quite popular tool in commutative algebra to represent graphs and to 
study their properties \cite{SVV,Vill}.  By using them, we generalize 
Reidys and Stadler's subgroup metric to define a metric through their 
permutation subgroups model, to define a family of metrics 
$(d_{m})_{m\geq 3}$ on the set of all contact structures of a fixed 
length.  Up to our knowledge, these are the first metrics defined on 
arbitrary contact structures of a fixed length that are independent of 
any notion of graph edition.  We express these metrics in terms of 
Hilbert functions, which makes them easily computable using several 
public domain computer algebra systems like for instance, 
CoCoA~\cite{cocoa} or Macaulay \cite{macaulay}.  We also obtain 
explicit expressions for several of these metrics on contact and RNA 
secondary structures, which allow to grasp the notion of similarity 
they measure.

We hope that our metrics will increase the range of sensible metrics 
available in the applications of the comparison of structures of a 
fixed length mentioned above: as Moulton, Zuker \textsl{et al}
point out, ``[\ldots] generally speaking, it is probably safest to 
try as many metrics as possible'' \cite[p.\ 290]{ZukM}.

\section{Preliminaries}

In this section we recall some definitions and facts on contact 
 and RNA secondary structures, and we take the opportunity 
to fix nome notations and conventions that we shall use henceforth, 
usually without any further notice.\smallskip

\noindent\textbf{\emph{Contact structures and RNA secondary 
structures}.} From now on, let $[n]$ denote the set $\{1,\ldots,n\}$, 
for every positive integer $n$.  We begin by recalling the definition 
of contact structure from \cite{RS96,SS99};  contact structures are also called 
\emph{diagrams} in \cite{HS99}.

\begin{definition}
A \emph{contact structure}  of length $n$ is an undirected graph 
without multiple edges or self-loops $\Gamma=([n],Q)$, for some $n\geqs 
1$, whose arcs $\{i,j\}\in Q$, called \emph{contacts}, satisfy the 
following condition:

i)  For every $i\in [n]$, $\{i,i+1\}\notin Q$.

\noindent A contact structure \emph{has unique bonds} when it satisfies the 
following extra condition:

ii) For every $i\in [n]$, if $\{i,j\},\{i,k\}\in Q$, then $j=k$.
\end{definition}

Condition (i) translates the impossibility of a contact between two 
consecutive monomers, while condition (ii) translates the {unique 
bonds condition} in RNA secondary structures mentioned in the 
introduction.  We shall call the contact structures with unique bonds 
\emph{RNA secondary structures}.  As we mentioned in the Introduction, 
the conventional definition of RNA secondary structure forbids moreover the 
existence of \emph{pseudoknots} (pairs of contacts $\{i,j\}$ 
and $\{k,l\}$ such that $i<k<j<l$), but we shall not impose this 
restriction here.

We shall denote from now on a contact $\{j,k\}$ by $j\!\cdot\!  k$ or 
$k\!\cdot\!  j$, without distinction.  A node is said to be 
\emph{isolated} in a contact structure when it is not involved 
in any contact.  

We shall often represent specific RNA secondary structures without 
pseudoknots by means of their \emph{bracket representation} 
\cite{Hof1}, obtained by replacing in the sequence $[n]$ each contact 
$i\!\cdot \!  j$ with $i<j$ by a ``('' in the $i$th position and a 
``)'' in the $j$th position, and each isolated node by a dot in the 
corresponding position.  For instance,
$$
((((((...)))))..((...)).)
$$
represents the secondary structure
$$
\Bigl([25],\{1\!\cdot\! 25, 2\!\cdot\!14, 
3\!\cdot\!13,4\!\cdot\!12,5\!\cdot\!11,6\!\cdot\!10,17\!\cdot\!23,18\!\cdot\!22\}\Bigr).
$$
Knotted RNA secondary structures admit a similar representation, using 
different types of brackets to represent contacts in order to avoid 
ambiguities.

Given two contact structures of the same length 
$\Gamma_{1}=([n],Q_{1}),\Gamma_{2}=([n],Q_{2})$, their \emph{union} is 
the contact structure
$$
\Gamma_{1}\cup\Gamma_{2}=([n],Q_{1}\cup Q_{2}).
$$

From now on, and unless otherwise stated, given any contact structure 
$\Gamma$ or $\Gamma_{i}$, $i=1,2,\ldots$, we shall always denote its 
set of contacts by $Q$ or $Q_{i}$, respectively.

Let $\Cn$ and $\SE$ denote the sets of all contact structures and of 
all RNA secondary structures of length $n$, respectively.  
\smallskip

\noindent\textbf{\emph{Subgroup metric}.} For every $n\geqs 1$, let 
$\Sn$ be the symmetric group of permutations of $[n]$.  In 
\cite{RS96}, Reidys and Stadler associated to every RNA secondary 
structure $\Gamma\in \SE$ the subgroup $G(\Gamma)$ of $\Sn$ generated 
by the set of the transpositions corresponding to the contacts in 
$\Gamma$:
$$
G(\Gamma)=\langle\{(i,j)\mid i\!\cdot \! j\in Q\}\rangle.
$$
They also proved that the mapping $\Gamma \mapsto 
G(\Gamma)$ is an embedding of $\SE$ into the 
set $\mbox{Sub}(\Sn)$ of subgroups of $\Sn$, and they used this 
representation of RNA secondary structures as permutation subgroups to 
define the following \emph{subgroup metric}:
$$
\begin{array}{rrcl}
d_{sgr}:&  \SE\times \SE & \to & \RR\\
& (\Gamma_{1},\Gamma_{2}) &\mapsto & 
\ln\left|\frac{\displaystyle G(\Gamma_{1})\cdot 
G(\Gamma_{2})}{\displaystyle G(\Gamma_{1})\cap G(\Gamma_{2})}\right|
\end{array}
$$
In \cite{Ros03} it was proved that this metric simply measures, up to 
a scalar factor, the cardinal $|Q_{1}\Delta Q_{2}|$ of the symmetric 
difference of the sets of contacts.  

Unfortunately, if we extend the mapping $G$ to the set $\Cn$ of all 
contact structures of length $n$, we no longer obtain an embedding 
into $\mbox{Sub}(\Sn)$, as the following easy example shows.

\begin{example}\label{ex-1}
Let $\Gamma_{1}=([5],Q_{1})$ and $\Gamma_{2}=([5],Q_{2})$ be contact 
structures with sets of contacts
$$
Q_{1}=\{1\!\cdot \! 3,3\!\cdot \! 5\},\
Q_{2}=\{1\!\cdot \! 5,3\!\cdot \! 5\}.
$$
see Fig.\ \ref{fig-ex1}.
Then $G(\Gamma_{1})=\langle (1,3), (3,5)\rangle$ and 
$G(\Gamma_{2})=\langle (1,5), (3,5)\rangle$ are both equal to
$$
\{\Id, (1,3), (1,5), (3,5), (1,3,5), (1,5,3)\}.
$$
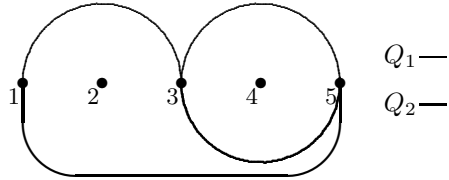
\begin{figure}[thb]
\begin{center}
\begin{picture}(170,50)(0,-10)
\multiput(0,20)(30,0){5}{\circle*{4}}
\put(-1,18){\makebox(0,0)[tr]{\small $1$}}
\put(29,18){\makebox(0,0)[tr]{\small $2$}}
\put(59,18){\makebox(0,0)[tr]{\small $3$}}
\put(89,18){\makebox(0,0)[tr]{\small $4$}}
\put(119,18){\makebox(0,0)[tr]{\small $5$}}
\thinlines
\put(30,20){\arc(30,2){180}}
\put(90,20){\arc(30,2){180}}
\thicklines
\put(90,20){\arc(-30,-2){180}}
\put(60,20){\oval(120,70)[b]}

\put(150,10){\begin{picture}(0,0)
   \put(-1,20){\makebox(0,0)[r]{$Q_{1}$}}
   \thinlines
   \put(0,20){\line(1,0){10}}
   \put(-1,2){\makebox(0,0)[r]{$Q_{2}$}}
   \thicklines
   \put(0,2){\line(1,0){10}}
   \end{picture}}
   \end{picture}
\end{center}
\caption{The contact structures in Example \ref{ex-1}.}
\label{fig-ex1}
\end{figure}
\end{example}

This entails in particular that the subgroup metric, when extended to 
the set $\Cn$, yields only a pseudodistance: it is nonnegative and 
symmetric and satisfies the triangular inequality, but 
$d_{sgr}(\Gamma_{1},\Gamma_{2})=0$ does not imply 
$\Gamma_{1}=\Gamma_{2}$.  The scope of the failure of the separability 
condition is determined by the following  result.

\begin{proposition}
For every $\Gamma_{1},\Gamma_{2}\in \Cn$, 
$d_{sgr}(\Gamma_{1},\Gamma_{2})=0$ if and only if for every $i\!\cdot 
\!  j\in \Gamma_{1}$ there exists a chain of contacts $k_{1} \!\cdot 
\!  k_{2}, k_{2} \!\cdot \!  k_{3},\ldots, k_{m-2} \!\cdot \!  k_{m-1}, 
k_{m-1} \!\cdot \!  k_{m}$ in $\Gamma_{2}$ with $m\geqs 2$, $k_{1}=i$ and 
$k_{m}=j$, and vice versa, for every $i\!\cdot \!  j\in \Gamma_{2}$ there is a 
similar chain of contacts in $\Gamma_{1}$ going from $i$ to $j$.
\end{proposition}

\begin{proof}
By \cite[Thm.\ 5]{RS96}, $d_{sgr}(\Gamma_{1},\Gamma_{2})=0$ if and 
only if $G(\Gamma_{1})=G(\Gamma_{2})$, i.e., if and only if every 
transposition corresponding to a contact in $\Gamma_{1}$ is a product 
of transpositions corresponding to contacts in $\Gamma_{2}$, and vice 
versa, every transposition corresponding to a contact in $\Gamma_{2}$ 
is a product of transpositions corresponding to contacts in 
$\Gamma_{1}$, a condition that is equivalent to the one given in the 
statement.
\end{proof}

\noindent\textbf{\emph{Orbits}.} Reidys and Stadler also represented an RNA 
secondary structure $\Gamma\in \SE$ with set of contacts
$Q=\{i_{1}\!\cdot\!j_{1},\ldots,i_{k}\!\cdot \!j_{k}\}$ by the 
involution\footnote{Notice that this product is only well-defined if the transpositions 
appearing in it commute with each other, and thus this definition does not make sense 
for arbitrary contact structures, at least unless some convention is introduced on 
the order how these transpositions must be composed; we shall not 
consider this problem here.}
$$
\pi(\Gamma)=\prod_{t=1}^k (i_{t},j_{t})\in \Sn.
$$
They also proved that this construction yields and embedding 
$\pi:\SE\hookrightarrow \Sn$, which they used to induce metrics on 
$\SE$ from metrics on $\Sn$ \cite{HS99,RS96}.

For every $\Gamma_{1},\Gamma_{2}\in \SE$, let 
$D(\Gamma_{1},\Gamma_{2}) =\langle 
\pi(\Gamma_{1}),\pi(\Gamma_{2})\rangle\in Sub(\Sn)$ be the dihedral 
subgroup of $S_{n}$ generated by the involutions associated to them.  This 
subgroup acts on $[n]$.  The \emph{orbits} induced by this action can 
be understood as subsets $\{i_{1},i_{2},\ldots,i_{m}\}\subseteq [n]$, 
$m\geqs 1$, such that
$$
i_{1}\!\cdot\!  i_{2},i_{2}\!\cdot\!  i_{3},\ldots,
i_{m-1}\!\cdot\!i_{m} \in Q_{1}\cup Q_{2}
$$
and maximal with this property, i.e., such that any other contact in 
$Q_{1}\cup Q_{2}$ involving some element of this subset can only be 
$i_{1}\!\cdot \!  i_{m}$.  Notice that these orbits are exactly the 
connected components of the graph $\Gamma_{1}\cup \Gamma_{2}$.  The 
unique bonds condition (or, in group-theoretical terms, the fact that 
$\pi(\Gamma_{1})$ and $\pi(\Gamma_{2})$ are involutions) implies that 
if $\{i_{1},i_{2},\ldots,i_{m}\}$ is such an orbit, then 
$i_{1}\!\cdot\!  i_{2},i_{3}\!\cdot\!  i_{4},\ldots$ belong to one of 
the sets $Q_{1}$ or $Q_{2}$ and 
$i_{2}\!\cdot\!i_{3},i_{4}\!\cdot\!i_{5},\ldots$ belong to the other 
one.  

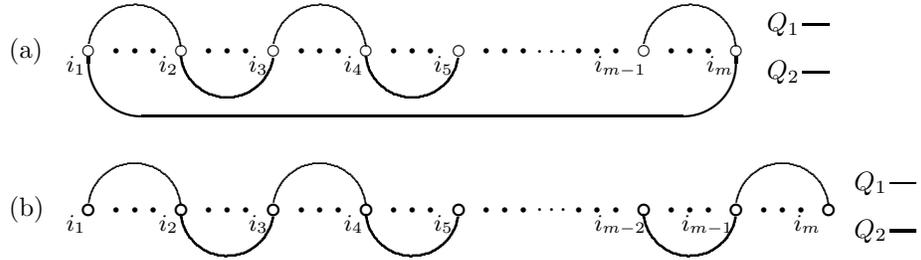
\begin{figure}[thb]
\begin{center}
\begin{picture}(320,100)(-20,30)
\multiput(0,110)(35,0){5}{\circle{4}}
\multiput(210,110)(35,0){2}{\circle{4}}
\multiput(10.5,110)(35,0){7}{\multiput(0,0)(7,0){3}{\circle*{2}}}
\thinlines
\put(176,110){\makebox(0,0){\dots}}
\put(-1,108){\makebox(0,0)[tr]{\small $i_{1}$}}
\put(34,108){\makebox(0,0)[tr]{\small $i_{2}$}}
\put(68,108){\makebox(0,0)[tr]{\small $i_{3}$}}
\put(104,108){\makebox(0,0)[tr]{\small $i_{4}$}}
\put(138,108){\makebox(0,0)[tr]{\small $i_{5}$}}
\put(211,108){\makebox(0,0)[tr]{\small $i_{m-1}$}}
\put(244,108){\makebox(0,0)[tr]{\small $i_{m}$}}

\thinlines
\put(17.5,110){\arc(17.5,2){164}}
\put(87.5,110){\arc(17.5,2){164}}
\put(227.5,110){\arc(17.5,2){164}}
\thicklines
\put(52.5,110){\arc(-17.5,-2){164}}
\put(122.5,110){\arc(-17.5,-2){164}}
\put(122.5,108){\oval(245,45)[b]}

\put(270,100){\begin{picture}(0,0)
   \put(-1,20){\makebox(0,0)[r]{$Q_{1}$}}
   \thinlines
   \put(0,20){\line(1,0){10}}
   \put(-1,2){\makebox(0,0)[r]{$Q_{2}$}}
   \thicklines
   \put(0,2){\line(1,0){10}}
   \end{picture}}
   \put(-30,110){\makebox(0,0)[l]{(a)}}
   
\multiput(0,50)(35,0){5}{\circle{4}}
\multiput(210,50)(35,0){3}{\circle{4}}

\multiput(10.5,50)(35,0){8}{\multiput(0,0)(7,0){3}{\circle*{2}}}
\thinlines
\put(176,50){\makebox(0,0){\dots}}
\put(-1,48){\makebox(0,0)[tr]{\small $i_{1}$}}
\put(34,48){\makebox(0,0)[tr]{\small $i_{2}$}}
\put(68,48){\makebox(0,0)[tr]{\small $i_{3}$}}
\put(104,48){\makebox(0,0)[tr]{\small $i_{4}$}}
\put(138,48){\makebox(0,0)[tr]{\small $i_{5}$}}
\put(211,48){\makebox(0,0)[tr]{\small $i_{m-2}$}}
\put(244,48){\makebox(0,0)[tr]{\small $i_{m-1}$}}
\put(277,48){\makebox(0,0)[tr]{\small $i_{m}$}}

\thinlines
\put(17.5,50){\arc(17.5,2){164}}
\put(87.5,50){\arc(17.5,2){164}}
\put(262.5,50){\arc(17.5,2){164}}
\thicklines
\put(52.5,50){\arc(-17.5,-2){164}}
\put(122.5,50){\arc(-17.5,-2){164}}
\put(227.5,50){\arc(-17.5,-2){164}}

\put(303,40){\begin{picture}(0,0)
   \put(-1,20){\makebox(0,0)[r]{$Q_{1}$}}
   \thinlines
   \put(0,20){\line(1,0){10}}
   \put(-1,2){\makebox(0,0)[r]{$Q_{2}$}}
   \thicklines
   \put(0,2){\line(1,0){10}}
   \end{picture}}
   \put(-30,50){\makebox(0,0)[l]{(b)}}
\end{picture}
\end{center}
\caption{A  cyclic orbit of length $m$ (a) and a
 linear orbit of length $m$ (b).}
\label{figorb}
\end{figure}

Such an orbit is \emph{cyclic} if $m=2$ and $i_{1}\!\cdot\!  
i_{2}\in Q_{1}\cap Q_{2}$, or $m\geqs 3$ and $i_{1}\!\cdot\!  i_{m}\in 
Q_{1}\cup Q_{2}$, and it is \emph{linear} in all other cases: see 
Fig.~\ref{figorb}.  We shall call the cardinal of an orbit its 
\emph{length}.  The length of a cyclic orbit is always even: if 
$i_{1}\!\cdot\!  i_{2}\in Q_{1}$ in a cyclic orbit 
$\{i_{1},i_{2},\ldots,i_{m}\}$, then $i_{1}\!\cdot\!  i_{m}\in Q_{2}$ 
and hence $i_{m-1}\!\cdot\!  i_{m}\in Q_{1}$.

 An orbit is \emph{trivial} when it is a singleton: it is a 
linear orbit consisting of a node that it is isolated in both 
$\Gamma_{1}$ and $\Gamma_{2}$. If $\{i_{1},i_{2},\ldots,i_{m}\}$ is a non-trivial linear orbit with 
$i_{1}\!\cdot\!  i_{2},i_{2}\!\cdot\!  i_{3},\ldots, 
i_{m-1}\!\cdot\!i_{m} \in Q_{1}\cup Q_{2}$ and $i_{1}\!\cdot\!  
i_{m}\notin Q_{1}\cup Q_{2}$, then $i_{1},i_{m}$ are its \emph{end 
points}. 

We shall say that a contact $i\!\cdot \!j\in Q_{1}\cup Q_{2}$ is 
\emph{involved} in an orbit when its vertices $i,j$ belong to this 
orbit.  Every contact in $Q_{1}\cup Q_{2}$ is involved in one and only 
one orbit, and a contact belongs to $Q_{1}\Delta Q_{2}$ if and only if 
it is involved in a linear orbit or in a cyclic orbit of length $m>2$.

Let, for every $k\geqs 2$,
$$
\Lambda^{(m)},\quad \Lambda_{\geq k},\quad \Theta^{(m)}
$$
denote, respectively, the number of linear orbits of length $m$, the 
number of linear orbits of length $m\geqs k$ and the number of cyclic 
orbits of length $m$ induced by the action of $D(\Gamma_{1},\Gamma_{2})$ 
on $[n]$.  Since a cyclic orbit of length $m$ involves $m$ contacts, 
and a linear orbit of length $m$ involves $m-1$ contacts, we have that
$$
|Q_{1}\Delta Q_{2}|=\sum_{m\geq 4}m\Theta^{(m)}+\sum_{m\geq 2}(m-1)\Lambda^{(m)}.
$$

\section{A family of metrics based on edge ideals}

Let $n$ be from now on an integer  greater than 2.  Let 
$\MM(x_{1},\ldots,x_{n})$, or simply
$\MM(\underline{x})$, be the set of all monomials in the variables 
$x_{1},\ldots,x_{n}$.  We shall denote a monomial 
$x_{1}^{\alpha_{1}}\cdots x_{n}^{\alpha_{n}}\in \MM(\underline{x})$ by 
$x^{(\alpha_{1},\ldots,\alpha_{n})}$ or simply by 
$x^{\underline{\alpha}}$ if we let $\underline{\alpha}$ stand for the 
$n$-tuple $(\alpha_{1},\ldots,\alpha_{n})$.  The \emph{total degree} 
of a monomial $x^{(\alpha_{1},\ldots,\alpha_{n})}$ is 
$\sum_{i=1}^n \alpha_{i}$.  
For every  $m\geqs 0$, 
\begin{itemize}
\item let
$\MM(\underline{x})^{(m)}$ be the set of all monomials in 
$\MM(\underline{x})$ of total degree $m$,
and

\item let $\MM(\underline{x})_{m}$ be the set of all monomials in 
$\MM(\underline{x})$ of total degree $\leqs m$.
\end{itemize}
Recall that 
$$
|\MM(\underline{x})_{m}|=\binom{n+m}{n}\mbox{ and } 
|\MM(\underline{x})^{(m)}|=\binom{n+m-1}{n-1}.
$$

Let $\FF$ be the field $\ZZ/2\ZZ$ and 
$\FF[x_{1},\ldots,x_{n}]$, or simply $\FF[\underline{x}]$, the ring of 
polynomials in the variables $x_{1},\ldots,x_{n}$ with coefficients in 
$\FF$.  Let $Id(\FF[\underline{x}])$ denote the set of ideals of 
$\FF[\underline{x}]$.  For every $I\in Id(\FF[\underline{x}])$ and for 
every $m\geqs 0$, 
\begin{itemize}
\item let $M(I)=I\cap \MM(\underline{x})$ be the set of all monomials that belong to $I$;
\item let $M(I)^{(m)}  = I\cap \MM(\underline{x})^{(m)}$ be the set of 
all monomials of total degree $m$ that belong to~$I$;
\item let
$M(I)_{m} =I\cap \MM(\underline{x})_{m}$ be the set of all monomials 
of total degree $\leqs m$ that belong to~$I$;
\item let $C(I)=\MM(\underline{x})-M(I)$
 be the set of all monomials that do not belong to~$I$; and
\item let
$C(I)_{m}=C(I)\cap\MM(\underline{x})_{m}$
 be the set of all monomials of total degree $\leqs m$ that do not belong 
 to~$I$.
 \end{itemize}

An ideal $I$ of $\FF[\underline{x}]$ is \emph{monomial} when it 
is generated by a set of monomials.  It should be recalled that, given 
a monomial ideal $I$ generated by a set of monomials $M$, the 
monomials in $M(I)$ are exactly those that are divisible by some 
monomial in $M$ and the polynomials in $I$ are exactly the linear 
combinations (with coefficients in $\FF$) of monomials in $M(I)$; in 
particular, for every two monomial ideals $I$ and $J$ of 
$\FF[\underline{x}]$, $I=J$ if and only if $M(I)=M(J)$.

\begin{definition}
For every $\Gamma=([n],Q)\in \Cn$, the \emph{edge ideal} $I_{\Gamma}$ 
of $\Gamma$ is the monomial ideal of $\FF[\underline{x}]$ generated by 
the products of pairs of variables whose indexes form a contact in 
$\Gamma$:
$$
I_{\Gamma}=\langle \{x_{i}x_{j}\mid i\!\cdot \! j\in Q\}\rangle.
$$
\end{definition}

\begin{proposition}\label{inj-1}
The mapping $I: \Cn \to \mbox{Id}(\FF[\underline{x}])$ that sends 
every $\Gamma\in \Cn$ to its edge ideal, is an embedding.
\end{proposition}

\begin{proof}
For every $\Gamma\in \Cn$, the monomials in $I_{\Gamma}$ are exactly those divisible by some 
$x_{i}x_{j}$ with $i\!\cdot \!  j\in Q$.  This implies that
$$
M(I_{\Gamma})^{(2)}=\{x_{i}x_{j}\mid i\!\cdot \!  j\in Q\},
$$
and therefore $\Gamma$ is uniquely determined by $M(I_{\Gamma})^{(2)}$.
\end{proof}

Given two contact structures $\Gamma_{1},\Gamma_{2}\in \Cn$, it is 
clear that
$$
I_{\Gamma_1}+I_{\Gamma_{2}}=\Bigl\langle\{x_{i}x_{j}\mid i\!\cdot\! j\in 
Q_{1}\}\cup \{x_{i}x_{j}\mid i\!\cdot\! j\in 
Q_{2}\}\Bigr\rangle=I_{\Gamma_{1}\cup \Gamma_{2}}.
$$
As far as $I_{\Gamma_{1}}\cap I_{\Gamma_{2}}$ goes,
it is straightforward to prove that it is generated by
$$
\begin{array}{l}
\{x_{i}x_{j}\mid i\!\cdot \! j\in Q_{1}\cap Q_{2}\} \cup 
\{x_{i}x_{j}x_{k}\mid i\!\cdot \! j\in 
Q_{1}-Q_{2},j\!\cdot \!k\in Q_{2}-Q_{1}\}\\
\qquad \cup \{x_{i}x_{j}x_{k}x_{l}\mid i\!\cdot \! j\in 
Q_{1}-Q_{2},k\!\cdot \!l\in Q_{2}-Q_{1},\{i,j\}\cap\{k,l\}\neq 
\emptyset\}.
\end{array}
$$


Using a construction similar to the one introduced by Reidys and 
Stadler for subgroups, we want to measure the difference between two 
contact structures $\Gamma_{1},\Gamma_{2}\in \Cn$ by means of the 
quotient $(I_{\Gamma_{1}}+I_{\Gamma_{2}})/(I_{\Gamma_{1}}\cap 
I_{\Gamma_{2}})$.  Notice that this quotient is a singleton if and 
only if $I_{\Gamma_{1}}=I_{\Gamma_{2}}$, i.e., if and only if 
$\Gamma_{1}=\Gamma_{2}$.  Unfortunately, in all other cases this 
quotient is infinite: if a monomial $x^{\underline{\alpha}}$ belongs 
to, say, $I_{\Gamma_{1}}-I_{\Gamma_{2}}$, then all its powers define 
pairwise different equivalence classes modulo $I_{\Gamma_{1}}\cap 
I_{\Gamma_{2}}$ Thus, to obtain a ``finite distance'' we move to 
quotients of $\FF[\underline{x}]$.

For every $n\geqs 1$ and $m\geqs 3$, let us consider the quotient ring 
$$
R_{n,m}=\FF[x_{1},\ldots,x_{n}]/\langle 
\MM(\underline{x})^{(m)}\rangle,
$$ 
and let $\pi_{m}:\FF[x_{1},\ldots,x_{n}]\to R_{n,m}$ be the corresponding 
quotient ring homomorphism.  For every $I\in Id(\FF[x_{1},\ldots,x_{n}])$, 
let $\pi_{m}(I)$ be the image of $I$ in $R_{n,m}$.

\begin{proposition}
For every $m\geqs 3$, the mapping
$d'_{m}:\Cn\times \Cn\to \RR$ defined by 
$$
d'_{m}(\Gamma_{1},\Gamma_{2})=\log_{2}\left| 
\frac{\pi_{m}(I_{\Gamma_{1}})+\pi_{m}(I_{\Gamma_{2}})} 
{\pi_{m}(I_{\Gamma_{1}})\cap \pi_{m}(I_{\Gamma_{2}})}\right|
$$
is a metric on $\Cn$.
\end{proposition}

\begin{proof}
When we perform the quotient $R_{n,m}=\FF[\underline{x}]/\langle 
\MM(\underline{x})^{(m)}\rangle$, all monomials with total degree 
greater or equal than $m$ are cancelled.  Then, each element in 
$R_{n,m}$ has a unique representative that is a linear combination 
with coefficients in $\FF$ of monomials of total degree at most $m-1$.  
Since $\FF$ is a finite field, this implies that $R_{n,m}$ is finite, 
and in particular it is a finite commutative group with the sum of 
quotient classes of polynomials as operation.  Let $Sub(R_{n,m})$ 
denote its set of subgroups.

On the other hand, since $m\geqs 3$, the quotient homomorphism $\pi_{m}$ 
does not identify any monomial of total degree 2 with any other 
monomial.  Thus, $\pi_{m}(I_{\Gamma_{1}})=\pi_{m}(I_{\Gamma_{2}})$ 
implies $M(I_{\Gamma_{1}})^{(2)}=M(I_{\Gamma_{2}})^{(2)}$ and hence, 
as we saw in Proposition \ref{inj-1}, 
$\Gamma_{1}=\Gamma_{2}$. In other words, the mapping $\pi_{m}\circ 
I:\Cn\to Sub(R_{n,m})$ sending every $\Gamma\in \Cn$ to
$\pi_{m}(I_{\Gamma})$, is an embedding.

Then, since by \cite[Thm.~5]{RS96} the mapping
$$
\Psi(I,J)=\log_{2}\left| \frac{I+J}{I\cap J}\right|,\quad I,J\in 
Sub(R_{n,m})
$$
is a metric on $Sub(R_{n,m})$, the mapping 
$$
d'_{m}(\Gamma_{1},\Gamma_{2})=\Psi(\pi_{m}(I_{\Gamma_{1}}),
\pi_{m}(I_{\Gamma_{2}})), \quad \Gamma_{1},\Gamma_{2}\in \Cn
$$
is a metric on $\Cn$, as we claimed.
\end{proof}

We have used $\log_{2}$ instead of $\ln$ in the definition of $d'_{m}$ 
in order to avoid unnecessary scalar factors: cf.\ \cite[Prop.\ 
4]{Ros03}.  

These metrics $d'_{m}$ have a simple description in terms of symmetric 
differences of sets of monomials.

\begin{proposition}\label{simdif}
For every $m\geqs 3$ and for every $\Gamma_{1},\Gamma_{2}\in \Cn$,
$$
d'_{m}(\Gamma_{1},\Gamma_{2}) = |M(I_{\Gamma_{1}})_{m-1}\Delta 
M(I_{\Gamma_{2}})_{m-1}|.
$$
\end{proposition}

\begin{proof}
Notice that, for every $I\in Id(\FF[\underline{x}])$,
$$
\pi_{m}(I)=\pi_{m}(I+\langle \MM(\underline{x})^{(m)}\rangle).
$$
Then, for every $\Gamma_{1},\Gamma_{2}\in \Cn$,
$$
\begin{array}{rl}
\frac{\displaystyle \pi_{m}(I_{\Gamma_{1}})+\pi_{m}(I_{\Gamma_{2}})}%
{\displaystyle \pi_{m}(I_{\Gamma_{1}})\cap \pi_{m}(I_{\Gamma_{2}})} & 
= \frac{\displaystyle\pi_{m}(I_{\Gamma_{1}}+\langle 
\MM(\underline{x})^{(m)}\rangle)+ \pi_{m}(I_{\Gamma_{2}}+\langle 
\MM(\underline{x})^{(m)}\rangle)} {\displaystyle 
\pi_{m}(I_{\Gamma_{1}}+\langle 
\MM(\underline{x})^{(m)}\rangle)\cap 
\pi_{m}(I_{\Gamma_{2}}+\langle 
\MM(\underline{x})^{(m)}\rangle)}\\[1ex] & = 
\frac{\displaystyle\pi_{m}\Bigl ((I_{\Gamma_{1}}+\MM(\underline{x})^{(m)}\rangle)+ 
(I_{\Gamma_{2}}+\langle 
\MM(\underline{x})^{(m)}\rangle)\Bigr)}{\displaystyle 
\pi_{m}\Bigl((I_{\Gamma_{1}}+\langle 
\MM(\underline{x})^{(m)}\rangle)\cap (I_{\Gamma_{2}}+\langle 
\MM(\underline{x})^{(m)}\rangle)\Bigr)} 
\\[1ex] & 
\cong  
\frac{\displaystyle (I_{\Gamma_{1}}+\langle 
\MM(\underline{x})^{(m)}\rangle)+
(I_{\Gamma_{2}}+\langle \MM(\underline{x})^{(m)}\rangle)} 
{\displaystyle (I_{\Gamma_{1}}+\langle \MM(\underline{x})^{(m)}\rangle)\cap 
(I_{\Gamma_{2}}+\langle \MM(\underline{x})^{(m)}\rangle)}\\[1ex]
& =
\frac{\displaystyle I_{\Gamma_{1}}+I_{\Gamma_{2}}+\langle 
\MM(\underline{x})^{(m)}\rangle} 
{\displaystyle (I_{\Gamma_{1}}\cap 
I_{\Gamma_{2}})+\langle \MM(\underline{x})^{(m)}\rangle},
\end{array}
$$
where the equality
$$
(I_{\Gamma_{1}}+\langle \MM(\underline{x})^{(m)}\rangle)\cap 
(I_{\Gamma_{2}}+\langle \MM(\underline{x})^{(m)}\rangle)=
(I_{\Gamma_{1}}\cap 
I_{\Gamma_{2}})+\langle \MM(\underline{x})^{(m)}\rangle
$$
used in the last step holds because $I_{\Gamma_{1}}$, $I_{\Gamma_{2}}$ 
and $\langle \MM(\underline{x})^{(m)}\rangle$ are monomial 
ideals.

To simplify the notations, set
$$
\begin{array}{rl}
J & = I_{\Gamma_{1}}+I_{\Gamma_{2}}+\langle 
\MM(\underline{x})^{(m)}\rangle,\\
K & =(I_{\Gamma_{1}}\cap I_{\Gamma_{2}})+\langle 
\MM(\underline{x})^{(m)}\rangle,
\end{array}
$$
so that
$$
\frac{\displaystyle \pi_{m}(I_{\Gamma_{1}})+\pi_{m}(I_{\Gamma_{2}})}%
{\displaystyle \pi_{m}(I_{\Gamma_{1}})\cap \pi_{m}(I_{\Gamma_{2}})} 
\cong
\frac{J}{K}.
$$
These ideals
$J$ and $K$ are also monomial and
$$
\begin{array}{rl}
M(J) & = M(I_{\Gamma_{1}})_{m-1}\cup M(I_{\Gamma_{2}})_{m-1} \cup 
\bigcup_{r>m-1} \MM(\underline{x})^{(r)},\\
 M(K) & = (M(I_{\Gamma_{1}})_{m-1}\cap 
M(I_{\Gamma_{2}})_{m-1})\cup \bigcup_{r>m-1} \MM(\underline{x})^{(r)}.
\end{array}
$$
A polynomial belongs to $J$ (resp.\ to $K$) if and only if it 
is a linear combination, with coefficients in $\FF$, of elements of 
$M(J)$ (resp.\ of $M(K)$).  This implies that every quotient class in 
$J/K$ has a unique representative of the form 
$\sum_{x^{\underline{\alpha}}\in M_{0}} x^{\underline{\alpha}}$ for 
some finite subset $M_{0}$ of $M(J)-M(K)$ (the zero class corresponds 
to $M_{0}=\emptyset$).
Since
$$
\begin{array}{rl}
M(J)-M(K) & =(M(I_{\Gamma_{1}})_{m-1}\cup M(I_{\Gamma_{2}})_{m-1}) - 
(M(I_{\Gamma_{1}})_{m-1}\cap M(I_{\Gamma_{2}})_{m-1}) \\ & =
M(I_{\Gamma_{1}})_{m-1}\Delta M(I_{\Gamma_{2}})_{m-1} 
\end{array}
$$
is a finite set, this implies that 
$$
\left|\frac{\displaystyle \pi_{m}(I_{\Gamma_{1}})+\pi_{m}(I_{\Gamma_{2}})}
{\displaystyle \pi_{m}(I_{\Gamma_{1}})\cap 
\pi_{m}(I_{\Gamma_{2}})}\right|  =
\left|\frac{J}{K}\right|=2^{|M(I_{\Gamma_{1}})_{m-1}\Delta 
M(I_{\Gamma_{2}})_{m-1}|},
$$
as we claimed.
\end{proof}

Proposition \ref{simdif} allows us to express the metrics $d'_{m}$ in 
terms of Hilbert functions.  For every monomial ideal $I$ of 
$\FF[\underline{x}]$ and for every $m\geqs 0$, let $H_{I}:\NN\to \NN$ 
be the mapping defined by
$$
H_{I}(m)=|C(I)_{m}|,\quad m\in \NN;
$$
i.e., $H_{I}(m)$ is the number of monomials of total degree $\leqs m$ 
that do not belong to $I$.  This mapping is called the (affine) 
\emph{Hilbert function} of $I$. It can be computed explicitly from 
a given finite set of generators of $I$ \cite{Big}; actually, several 
freely available 
computer algebra systems like, for instance, CoCoA~\cite{cocoa} or 
Macaulay \cite{macaulay}, compute Hilbert functions.

For every contact structure $\Gamma\in \Cn$, let $H_{\Gamma}$ denote 
the Hilbert function of its edge ideal.

\begin{corollary}\label{cor-Hm->dm}
For every $\Gamma_{1},\Gamma_{2}\in \Cn$ and for every $m\geqs 3$,
$$
d'_{m}(\Gamma_{1},\Gamma_{2})=H_{{\Gamma_{1}}}(m-1)+H_{{\Gamma_{2}}}(m-1)-
2H_{{\Gamma_{1}}\cup\Gamma_{2}}(m-1).
$$
\end{corollary}

\begin{proof}
We have that
$$
\begin{array}{l}
M(I_{\Gamma_{1}})_{m-1}\Delta 
M(I_{\Gamma_{2}})_{m-1}\\
\quad =(M(I_{\Gamma_{1}})_{m-1}-(M(I_{\Gamma_{1}})_{m-1}\cap 
M(I_{\Gamma_{2}})_{m-1}))\cup 
(M(I_{\Gamma_{2}})_{m-1}-(M(I_{\Gamma_{1}})_{m-1}\cap 
M(I_{\Gamma_{2}})_{m-1}))\\
\quad =(M(I_{\Gamma_{1}}+I_{\Gamma_{2}})_{m-1}-M(I_{\Gamma_{2}})_{m-1})\cup 
(M(I_{\Gamma_{1}}+I_{\Gamma_{2}})_{m-1}-M(I_{\Gamma_{1}})_{m-1})\\
\quad =( C(I_{\Gamma_{2}})_{m-1}-C(I_{\Gamma_{1}}+I_{\Gamma_{2}})_{m-1}) 
\cup (C(I_{\Gamma_{1}})_{m-1}-C(I_{\Gamma_{1}}+I_{\Gamma_{2}})_{m-1})
\end{array}
$$
and thus, this union being disjoint,
$$
\begin{array}{l}
|M(I_{\Gamma_{1}})_{m-1}\Delta M(I_{\Gamma_{2}})_{m-1}|\\
\qquad\quad = 
(|C(I_{\Gamma_{2}})_{m-1}|-|C(I_{\Gamma_{1}}+I_{\Gamma_{2}})_{m-1}|) 
+ (|C(I_{\Gamma_{1}})_{m-1}|-|C(I_{\Gamma_{1}}+I_{\Gamma_{2}})_{m-1}|)
\\
\qquad\quad = 
(H_{\Gamma_{2}}(m-1)-H_{\Gamma_{1}\cup\Gamma_{2}}(m-1))+
(H_{\Gamma_{1}}(m-1)-H_{\Gamma_{1}\cup\Gamma_{2}}(m-1)),
\end{array}
$$
as we claimed.
\end{proof}

To close this section, we want to point out that the metrics $d'_{m}$ 
grow with $n$ and $m$, and thus it is convenient to normalize them in 
order to avoid unnecessarily high figures.  More specifically, let 
$\Gamma_{0}=([n],\emptyset)$ be the \emph{empty RNA secondary 
structure} of length $n$ and let $\Gamma_{1}$ be an RNA secondary 
structure of length $n$ with only one contact, say $i\!\cdot \!  j$ 
with $i<j$.  Then $I_{\Gamma_{0}}= \{0\}$, 
$I_{\Gamma_{1}}=I_{\Gamma_{0}}+ I_{\Gamma_{1}}=\langle 
x_{i}x_{j}\rangle$, and therefore, for every $m\geqs 3$,
$$
d'_{m}(\Gamma_{0},\Gamma_{1})=H_{{\Gamma_{0}}}(m-1)-H_{{\Gamma_{1}}}(m-1),
$$
where
$$
\begin{array}{rl}
H_{{\Gamma_{0}}}(m-1) & =|\MM(\underline{x})_{m-1}|= 
\binom{n+m-1}{n}\\[1ex]
H_{{\Gamma_{1}}}(m-1)
& =|\MM(x_{1},\ldots,x_{i-1},x_{i+1},\ldots,x_{n})_{m-1}| \\
&\qquad +|\MM(x_{1},\ldots,x_{j-1},x_{j+1},\ldots,x_{n})_{m-1}|\\
& \qquad -|\MM(x_{1},\ldots,x_{i-1},x_{i+1},\ldots, 
x_{j-1},x_{j+1},\ldots,x_{n})_{m-1}|\\
& \!\!\!=2\binom{n+m-2}{n-1}-\binom{n+m-3}{n-2},
\end{array}
$$
and hence
$$
d'_{m}(\Gamma_{0},\Gamma_{1})=\binom{n+m-1}{n}-2\binom{n+m-2}{n-1}+
\binom{n+m-3}{n-2}=\binom{n+m-3}{n}.
$$
If we take 1 as the ``natural'' value for the distance between 
$\Gamma_{0}$ and $\Gamma_{1}$, then instead of using the metrics 
$d_{m}'$ on $\Cn$, we must divide them by $\binom{n+m-3}{n}$.

\begin{definition}
For every $m\geqs 3$, the \emph{edge ideal $m$th metric} on $\Cn$ is 
$$
d_{m}(\Gamma_{1},\Gamma_{2})=\frac{1}{\binom{n+m-3}{n}}
d'_{m}(\Gamma_{1},\Gamma_{2}),\quad \Gamma_{1},\Gamma_{2}\in \Cn.
$$
\end{definition}

So, for instance, on $\Cn$
$$
d_{3}=d'_{3},\quad
d_{4}=\frac{1}{n+1}d'_{4},\quad
d_{5}=\frac{1}{\binom{n+2}{2}}d'_{5},\mbox{ and so on}.
$$ 

Even after this modification, the metric $d_{m}$ is sensitive to $n$, 
in the sense that if we add to two contact structures of a given 
length $n$ an isolated point, making them contact structures of length 
$n+1$, then their distance $d_{m}$ (for $m\geq 4$: see Proposition 
\ref{prop-d3} below) may grow.  For instance, let $\Gamma_{0}$ be  
again the empty RNA secondary structure of length $n\geq 6$ and let now 
$\Gamma_{1}=([n],\{1\!\cdot\!3,4\!\cdot 6\})$.  Then
$$
d_{m}(\Gamma_{0},\Gamma_{1})=\frac{1}{\binom{n+m-3}{n}}
(H_{{\Gamma_{0}}}(m-1)-H_{{\Gamma_{1}}}(m-1)).
$$
We have seen above that $H_{{\Gamma_{0}}}(m-1)=\binom{n+m-1}{n}$, and 
we shall see in Proposition \ref{H-RNA} below that (with the convention 
that $\binom{i}{n}=0$ if $i<n$)
$$
H_{{\Gamma_{1}}}(m-1)=\binom{n+m-1}{n}-2\binom{n+m-3}{n}+\binom{n+m-5}{n}.
$$
Therefore, 
$$
\begin{array}{rl}
d_{m}(\Gamma_{0},\Gamma_{1}) & =\frac{1}{\binom{n+m-3}{n}}
(2\binom{n+m-5}{n}-\binom{n+m-3}{n})\\
& =2-\binom{n+m-3}{n}\Big/\binom{n+m-5}{n}
=2-\frac{(m-3)(m-4)}{(n+m-3)(n+m-4)}
\end{array}
$$
which increases with $n$ if $m\geq 5$. 

Since we are only interested in comparing contact structures of the 
same length, this sensitiveness of the edge ideal metrics to the 
length $n$ is not a major drawback.

\section{Some computations}

In this section we shall compute explicitly  some edge ideal $m$th
metrics on $\Cn$ and $\SE$, for low values of $m$. We begin with $m=3$.

\begin{proposition}\label{prop-d3}
For every $\Gamma_{1},\Gamma_{2}\in \Cn$, 
$$
d_{3}(\Gamma_{1},\Gamma_{2})=|Q_{1}\Delta Q_{2}|.
$$
\end{proposition}

\begin{proof}
Notice that $M(I_{\Gamma})_{1}=\emptyset$ for every $\Gamma\in \Cn$. 
Therefore
$$
M(I_{\Gamma_{1}})_{2}\Delta M(I_{\Gamma_{2}})_{2} 
=M(I_{\Gamma_{1}})^{(2)}\Delta M(I_{\Gamma_{2}})^{(2)}
=\{x_{i}x_{j}\mid 
i\!\cdot \!j\in (Q_{1}-Q_{2})\cup (Q_{2}-Q_{1})\}
$$
and hence $|M(I_{\Gamma_{1}})_{2}\Delta 
M(I_{\Gamma_{2}})_{2}|=|Q_{1}\Delta Q_{2}|$.
\end{proof}

Actually, it is not difficult to prove that, for every $\Gamma\in 
\SE$, the mapping $G(\Gamma)\to \pi_{3}(I_{\Gamma})$ sending every 
permutation $\sigma=(i_{1},j_{1})\cdots (i_{l},j_{l})\in G(\Gamma)$, 
with $i_{1}\!\cdot\!  j_{1},\ldots,i_{l}\!\cdot\!  j_{l}\in Q$, to the 
equivalence class of the polynomial $x_{i_{1}}x_{j_{1}}+\cdots 
+x_{i_{l}}x_{j_{l}}\in I_{\Gamma}$ modulo 
$\langle\MM(\underline{x})^{(m)}\rangle$, is an isomorphism of groups, 
considering $\pi_{3}(I_{\Gamma})$ as a subgroup of $R_{n,3}$.  This is 
not true for arbitrary contact structures, because in this case 
$G(\Gamma)$ need not be commutative, while $\pi_{3}(I_{\Gamma})$ is 
always so.  Therefore, the embedding $\pi_{3}\circ 
I:\Cn\hookrightarrow Sub(R_{n,3})$ generalizes the embedding 
$G:\SE\hookrightarrow Sub(\Sn)$, and hence the metric $d_{3}$ 
generalizes (up to a scalar factor) the subgroup metric $d_{sgr}$ at a 
level deeper than their raw value.

The edge ideal $m$th metrics for $m>3$ have a much more involved 
expression.  In their computation we shall use the following lemma; 
notice that the edge ideals of contact structures are radical monomial 
proper (i.e., $\neq \FF[\underline{x}]$) ideals.

\begin{lemma}\label{lema-N}
Let $I$ be a radical monomial proper ideal of $\FF[\underline{x}]$ and, for 
every $k\geqs 1$, let $SF_{k}(I)$ be the number of square free 
monomials of total degree $k$ belonging to $M(I)$. Then, for every $m\geqs 0$,
$$
H_{I}(m)= \binom{n+m}{n}- \sum_{k=1}^{m} \binom{m}{k}SF_{k}(I).
$$
\end{lemma}

\begin{proof}
If $I$ is a radical monomial ideal, then a monomial of the form 
$x_{i_{1}}^{\alpha_{i_{1}}}\cdots x_{i_{k}}^{\alpha_{i_{k}}}$, with 
$i_{1},\ldots,i_{k}$ pairwise different and each $\alpha_{i_{t}}\geqs 
1$, belongs to $M(I)$ if and only if the corresponding square free monomial 
$x_{i_{1}}\cdots x_{i_{k}}$ belongs to $M(I)$.  
Therefore, each one of the $SF_{k}(I)$ square free monomials 
$x_{i_{1}}\cdots x_{i_{k}}$ of total degree $k\geqs 1$ in $M(I)$ adds 
as many monomials $x_{i_{1}}^{\alpha_{i_{1}}}\cdots 
x_{i_{k}}^{\alpha_{i_{k}}}$ to $M(I)_{m}$ as vectors 
$(\alpha_{i_{1}},\ldots,\alpha_{i_{k}})\in (\NN-\{0\})^k$ such that 
$\sum_{t=1}^{k}\alpha_{i_{t}}\leqs m$ there exist, and the number of 
the latter is $\binom{k+m-k}{k}=\binom{m}{k}$.  Since all monomials in $M(I)_{m}$ added 
in this way are pairwise different and $1\notin I$ by assumption, this proves that
$$
|M(I)_{m}|= \sum_{k=1}^{m} \binom{m}{k}SF_{k}(I),
$$
and hence
$$
|C(I)_{m}|=|\MM(\underline{x})_{m}|-|M(I)_{m}|=\binom{n+m}{n}-
\sum_{k=1}^{m} \binom{m}{k}SF_{k}(I),
$$
as we claimed.
\end{proof}

Notice that if $\Gamma\in\Cn$, then $SF_{1}(I_{\Gamma})=0$ and 
$SF_{2}(I_{\Gamma})=|Q|$.

Let us compute now $d_{4}$ on $\Cn$.  For every contact structure 
$\Gamma\in \Cn$, let
$$
\begin{array}{rl}
A(\Gamma) & =\Bigl|\Bigl\{\{i\!\cdot\! j, j\!\cdot\! k\}\subseteq Q\mid j\neq k\Bigr\}\Bigr|\\
T(\Gamma) & =\Bigl|\Bigl\{\{i,j,k\}\subseteq [n]\mid i\!\cdot\! j, j\!\cdot\! 
k,i\!\cdot \! k\in Q\Bigr\}\Bigr|
\end{array}
$$
In other words, $A(\Gamma)$ and $T(\Gamma)$ are respectively the 
numbers of \emph{angles} and \emph{triangles}  
in $\Gamma$. Notice that each triangle contains three different angles
and therefore $3T(\Gamma)\leqs A(\Gamma)$.

\begin{proposition}\label{prop-d4}
For every $\Gamma_{1},\Gamma_{2}\in \Cn$,
$$
\begin{array}{rl}
d_{4}(\Gamma_{1},\Gamma_{2})=&
|Q_{1}\Delta Q_{2}|\\[1ex]
&\quad \displaystyle-\frac{1}{n+1}\Bigl(2A(\Gamma_{1}\cup 
\Gamma_{2})-A(\Gamma_{1})-A(\Gamma_{2})+2T(\Gamma_{1}\cup 
\Gamma_{2})-T(\Gamma_{1})-T(\Gamma_{2})\Bigr)
\end{array}
$$
\end{proposition}

\begin{proof}
For every $\Gamma=([n],Q)\in \Cn$ we have that
$$
H_{\Gamma}(3)=\binom{n+3}{n}-3SF_{1}(I_{\Gamma})-3SF_{2}(I_{\Gamma})
-SF_{3}(I_{\Gamma}),
$$
where $SF_{1}(I_{\Gamma})=0$ and 
$SF_{2}(I_{\Gamma})=|Q|$. It remains to compute $SF_{3}(I_{\Gamma})$:

\begin{itemize}
\item[(1)] For every $i\!\cdot \!j\in Q$, there are $(n-2)$ square free 
monomials $x_{i}x_{j}x_{k}$ in $M(I_{\Gamma})$: this makes $(n-2)|Q|$ 
such monomials.

\item[(2)] Now, if $i\!\cdot \!j,j\!\cdot \!k\in Q$ form an angle, the 
monomial $x_{i}x_{j}x_{k}$ was counted twice in (1): therefore, to 
count these monomials only once, we must subtract $A(\Gamma)$.

\item[(3)] Finally, if the nodes $i,j,k$ form a triangle in $\Gamma$, 
then the monomial $x_{i}x_{j}x_{k}$ was counted three times in (1) and 
it was subtracted three times in (2); therefore, to retrieve these 
monomials, we must add $T(\Gamma)$ 
again.
\end{itemize}

Therefore
$$
SF_{3}(I_{\Gamma})=(n-2)|Q|-A(\Gamma)+T(\Gamma)
$$
and
$$
H_{\Gamma}(3)=\binom{n+3}{n} -(n+1)|Q|+A(\Gamma)-T(\Gamma).
$$
We have then
$$
\begin{array}{l}
d_{4}'(\Gamma_{1},\Gamma_{2})= H_{{\Gamma_{1}}}(3)+H_{{\Gamma_{2}}}(3)- 
2H_{{\Gamma_{1}}\cup\Gamma_{2}}(3)\\
\qquad 
=\binom{n+3}{n} -(n+1)|Q_{1}|+A(\Gamma_{1})-T(\Gamma_{1})+\binom{n+3}{n} 
-(n+1)|Q_{2}|+A(\Gamma_{2})-T(\Gamma_{2})\\
\qquad\qquad -2\Bigl(\binom{n+3}{n} -(n+1)|Q_{1}\cup 
Q_{2}|+A(\Gamma_{1}\cup\Gamma_{2})-T(\Gamma_{1}\cup\Gamma_{2})\Bigr)\\
\qquad = (n+1)|Q_{1}\Delta Q_{2}|- (2A(\Gamma_{1}\cup 
\Gamma_{2})-A(\Gamma_{1})-A(\Gamma_{2})+2T(\Gamma_{1}\cup 
\Gamma_{2})-T(\Gamma_{1})-T(\Gamma_{2}));
\end{array}
$$
in the last equality we have used that $2|Q_{1}\cup Q_{2}|-|Q_{1}|-|Q_{2}|=|Q_{1}\Delta 
Q_{2}|$.

Dividing by $n+1$ this last expression for $d_{4}'(\Gamma_{1},\Gamma_{2})$, we obtain the expression for 
$d_{4}(\Gamma_{1},\Gamma_{2})$ given in the statement.
\end{proof}

A simple computation shows that, for every $\Gamma_{1},\Gamma_{2}\in 
\Cn$,
$$
\begin{array}{l}
2A(\Gamma_{1}\cup \Gamma_{2})-A(\Gamma_{1})-A(\Gamma_{2})\\
\qquad =\Bigl|\Bigl\{\{i\!\cdot\!j,j\!\cdot\! k\}\mid i\neq k,\
(i\!\cdot\!j,j\!\cdot\! k\in Q_{s})\mbox{ and }(i\!\cdot\!j \mbox{ or }
j\!\cdot\! k\notin Q_{t}),\mbox{ for some $\{s,t\}=\{1,2\}$}\Bigr\}\Bigr|\\
\qquad\qquad +2
\Bigl|\Bigl\{\{i\!\cdot\!j,j\!\cdot\! k\}\mid 
i\!\cdot\!j\in Q_{1}-Q_{2}, j\!\cdot\! k\in Q_{2}-Q_{1}\Bigr\}\Bigr|\\[2ex]
2T(\Gamma_{1}\cup \Gamma_{2})-T(\Gamma_{1})-T(\Gamma_{2})\\
\qquad=\Bigl|\Bigl\{\{i,j, k\}\mid 
(i\!\cdot\!j,j\!\cdot\! k,i\!\cdot\!k\in Q_{s})\mbox{ and 
}(i\!\cdot\!j, j\!\cdot\!k \mbox{ or 
}i\!\cdot\! k\notin Q_{t}),\mbox{ for some $\{s,t\}=\{1,2\}$}\Bigr\}\Bigr|\\
\qquad\qquad+2
\Bigl|\Bigl\{\{i,j, k\}\mid 
i\!\cdot\!j,j\!\cdot\! k\in Q_{s}\mbox{ and 
}i\!\cdot\! k\in Q_{t}-Q_{s},\mbox{ for some 
$\{s,t\}=\{1,2\}$}\Bigr\}\Bigr|
\end{array}
$$

\begin{example}
Let $\Gamma_{0}=([9],\{1\!\cdot\!3,4\!\cdot\!6\})$, and consider the 
following ``modifications'' of it:
$$
\begin{array}{c}
\Gamma_{1}  =([9],\{1\!\cdot\!3,4\!\cdot\!6, 7\!\cdot \!9\}),\ 
\Gamma_{2} =([9],\{1\!\cdot\!3,4\!\cdot\!6, 6\!\cdot \!9\}),\
\Gamma_{3}  =([9],\{1\!\cdot\!3,4\!\cdot\!6, 1\!\cdot \!6\})\\
\Gamma_{4}  =([9],\{1\!\cdot\!3,4\!\cdot\!7\}),\
\Gamma_{5}  =([9],\{1\!\cdot\!3,3\!\cdot\!6\}),\
\Gamma_{6}  =([9],\{1\!\cdot\!3,3\!\cdot\!5\}),\ 
\Gamma_{7}  =([9],\{1\!\cdot\!3,5\!\cdot\!7\})
\end{array}
$$
The contact structures $\Gamma_{1}$, $\Gamma_{2}$ and $\Gamma_{3}$ are 
obtained by adding a contact to $\Gamma_{0}$ in three different ways, 
$\Gamma_{4}$ and $\Gamma_{5}$ are obtained by shifting the contact 
$4\!\cdot\!6$ in two different ways, and $\Gamma_{6}$ and $\Gamma_{7}$ 
are obtained by displacing this contact in two more ways.  Notice that 
$\Gamma_{0},\Gamma_{1},\Gamma_{4}$ and $\Gamma_{7}$ are RNA secondary 
structures, but not the others.

We have that
$$
\begin{array}{c}
d_{3}(\Gamma_{0},\Gamma_{1})=d_{3}(\Gamma_{0},\Gamma_{2})=d_{3}(\Gamma_{0},\Gamma_{3})=1,\\
d_{3}(\Gamma_{0},\Gamma_{4})=d_{3}(\Gamma_{0},\Gamma_{5})=
d_{3}(\Gamma_{0},\Gamma_{6})=d_{3}(\Gamma_{0},\Gamma_{7})=2,
\end{array}
$$
while
$$
\begin{array}{c}
d_{4}(\Gamma_{0},\Gamma_{1})=1,\quad
d_{4}(\Gamma_{0},\Gamma_{2})=0.9,\quad
d_{4}(\Gamma_{0},\Gamma_{3})=0.8,\\
d_{4}(\Gamma_{0},\Gamma_{4})=1.8,\quad
d_{4}(\Gamma_{0},\Gamma_{5})=1.7,\quad
d_{4}(\Gamma_{0},\Gamma_{6})=1.9,\quad
d_{4}(\Gamma_{0},\Gamma_{7})=2
\end{array}
$$
\end{example}

The expression for $d_{4}$ on RNA secondary structures is much  
simpler.  Recall that, for every $\Gamma_{1},\Gamma_{2}\in \SE$, 
$\Lambda_{\geq 2}$ stands for the number of linear orbits of length 
$\geqs 2$, i.e., of non-trivial linear orbits induced by the action 
of $D(\Gamma_{1},\Gamma_{2})$ on $[n]$.

\begin{proposition}\label{prop-d4-RNA}
For every $\Gamma_{1},\Gamma_{2}\in \SE$, 
$$
\begin{array}{rl}
d_{4}(\Gamma_{1},\Gamma_{2})& =|Q_{1}\Delta 
Q_{2}|-\frac{2}{n+1}A(\Gamma_{1}\cup \Gamma_{2}) \\[2ex]
& =|Q_{1}\Delta 
Q_{2}|-\frac{2}{n+1}(|Q_{1}\Delta Q_{2}|- \Lambda_{\geq 2}).
\end{array}
$$
\end{proposition}

\begin{proof}
Notice that the unique bonds condition implies in this case that
$$
A(\Gamma_{1})=A(\Gamma_{2})=T(\Gamma_{1})=T(\Gamma_{2}) = 
T(\Gamma_{1}\cup \Gamma_{2})=0,
$$
from which the first equality follows. It remains to prove that if 
$\Gamma_{1},\Gamma_{2}\in SE$, then
$$
A(\Gamma_{1}\cup \Gamma_{2})=|Q_{1}\Delta Q_{2}|- 
\Lambda_{\geq 2}.
$$
To prove it, notice that if $\{i\!\cdot\!j,j\!\cdot\!  k\}$ forms an 
angle in $\Gamma_{1}\cup\Gamma_{2}$, then, again by the unique bonds 
condition, one these contacts must belong to $Q_{1}-Q_{2}$ and the 
other one to $Q_{2}-Q_{1}$, and the nodes $i,j,k$ belong to the same 
orbit of length at least 3.  Now, each cyclic orbit of length 
$m>2$ contains $m$ such pairs of contacts, while any linear orbit of length 
$m\geqs 2$ contains $m-2$ such pairs.
Then
$$
\begin{array}{rl}
A(\Gamma_{1}\cup \Gamma_{2}) & =\sum_{m\geq 4}m\Theta^{(m)}+\sum_{m\geq 
2}(m-2)\Lambda^{(m)}\\ & =\sum_{m\geq 4}m\Theta^{(m)}+\sum_{m\geq 
2}(m-1)\Lambda^{(m)} -\sum_{m\geq 2}\Lambda^{(m)} = |Q_{1}\Delta 
Q_{2}|- \Lambda_{\geq 2},
\end{array}
$$
as we wanted to prove.
\end{proof}

Therefore, on $\SE$, the metric $d_{4}$ increases with the cardinal of 
$Q_{1}\Delta Q_{2}$, but decreases with the number of pairs of 
contacts in $Q_{1}\Delta Q_{2}$ that share a node.  Notice moreover 
that
$$
0\leqs |Q_{1}\Delta Q_{2}|-\Lambda_{\geq 2}\leqs |Q_{1}\Delta Q_{2}|;
$$
the lower bound is achieved when all non-trivial orbits are linear of 
length 2 (i.e., when $\Gamma_{1}\cup\Gamma_{2}$ is again an RNA 
secondary structure), and the upper bound when all non-trivial orbits 
are cyclic (i.e., when $\Gamma_{1}$ and $\Gamma_{2}$ have exactly the 
same isolated nodes).

\begin{example}
According to \cite[\S 5.3.2]{SS99}, most RNA AU-sequences of 
length 16 do not fold at all, i.e., they form the empty RNA secondary structure 
$\Gamma_{0}=([16],\emptyset)$.  But some of these sequences (about a 3\% 
of them) do fold, forming one of the following three RNA secondary 
structures without pseudoknots:
$$
\Gamma_{1}:  .((((((...))))))\qquad
\Gamma_{2}:  ((((((...)))))).\qquad
\Gamma_{3}:  ((((((....))))))
$$
Recall that these are the bracket representations of
$$
\begin{array}{rl}
\Gamma_{1} & =\Bigl([16], \{2\!\cdot\! 16,3\!\cdot\! 15,4\!\cdot\! 14,5\!\cdot\! 
13,6\!\cdot\! 12, 7\!\cdot\! 11\}\Bigr)\\
\Gamma_{2} & =\Bigl([16], \{1\!\cdot\! 15,2\!\cdot\! 14,3\!\cdot\! 
13,4\!\cdot\! 12, 5\!\cdot\! 11,  6\!\cdot\! 10\}\Bigr)\\
\Gamma_{3} & =\Bigl([16], \{1\!\cdot\! 16,2\!\cdot\! 15,3\!\cdot\! 14,4\!\cdot\! 
13,5\!\cdot\! 12, 6\!\cdot\! 11\}\Bigr)
\end{array}
$$

They have pairwise disjoint sets of contacts, and hence
$$
d_{3}(\Gamma_{1},\Gamma_{2})=d_{3}(\Gamma_{1},\Gamma_{3})=
d_{3}(\Gamma_{2},\Gamma_{3})=12.
$$
But 
$$
d_{4}(\Gamma_{1},\Gamma_{2})=\frac{184}{17},\
d_{4}(\Gamma_{1},\Gamma_{3})=
d_{4}(\Gamma_{2},\Gamma_{3})=\frac{182}{17},
$$
which shows that, under $d_{4}$, $\Gamma_{1}$ and $\Gamma_{2}$ are 
closer to $\Gamma_{3}$ than to each other.
\end{example}

\begin{example}\label{ex-hairp}
Let us consider a hairpin with an interior loop $\Gamma_{0}\in 
\mathcal{S}_{21}$ with bracket representation
$$
((.((((...)))).....))
$$
One possible rearrangement of this secondary structure splits the 
hairpin into a multibranched loop by means of two shift moves, 
yielding a multibranched structure:
$$
\Gamma_{0}: ((.((((...)))).....))\Longrightarrow \Gamma_{1}:
((..(((...)))(...).))\Longrightarrow \Gamma_{2}: ((...((...))((...))))
$$
Another possible rearrangement through shift moves widens the final 
loop of the hairpin by moving a one-nucleotide bulge:
$$
\begin{array}{rl}
\Gamma_{0}: ((.((((...)))).....))& \Longrightarrow \Gamma'_{1}: 
(((.(((...)))).....))\Longrightarrow \Gamma'_{2}: 
((((.((...)))).....))\\ & \Longrightarrow \Gamma'_{3}: 
(((((.(...)))).....)) \Longrightarrow \Gamma'_{4}: 
((((((....)))).....))
\end{array}
$$
Now notice that
$$
d_{3}(\Gamma_{0},\Gamma_{1})=d_{3}(\Gamma_{0},\Gamma'_{1})=2,\
d_{3}(\Gamma_{0},\Gamma_{2})=d_{3}(\Gamma_{0},\Gamma'_{2})=4.
$$
But, although
$$
d_{4}(\Gamma_{0},\Gamma_{1})=d_{4}(\Gamma_{0},\Gamma'_{1})=\frac{21}{11},
$$
it turns out that
$$
d_{4}(\Gamma_{0},\Gamma_{2})=\frac{42}{11},\quad d_{4}(\Gamma_{0},\Gamma'_{2})=\frac{41}{11}.
$$
Therefore, under $d_{4}$, $\Gamma_{2}'$ is closer to $\Gamma_{0}$ than 
$\Gamma_{2}$. 
\end{example}

As $m$ grows, the description of $d_{m}$ on $\Cn$ gets more and more 
involved or, if we want it to remain simple, more and more 
uninformative.  The same happens on $\SE$, but at a lower pace.  
Therefore, from now on, we shall only consider edge metrics on RNA secondary 
structures. 

We have a closed formula for the Hilbert function of the edge ideal 
of an RNA secondary structure, given by the following result, which 
we consider interesting in itself.  In it 
we use the convention that $\binom{0}{0}=1$ and $\binom{0}{j}=0$ if 
$j>0$.

\begin{proposition}\label{H-RNA}
For every $\Gamma=([n],Q)\in \SE$ and for every $m\geqs 0$,
$$
H_{{\Gamma}}(m)=\sum_{j=0}^{\lfloor m/2\rfloor} (-1)^j\binom{|Q|}{j}
\binom{n+m-2j}{n}.
$$
\end{proposition}

\begin{proof}
To begin with, notice that the Hilbert function $H_{{\Gamma}}$ only 
depends on $|Q|$ because, for every two RNA secondary structures 
$\Gamma_{1},\Gamma_{2}$ with the same number of contacts, their edge 
ideals are the same up to a permutation of the variables and thus 
$|C(I_{\Gamma_{1}})_{m}|=|C(I_{\Gamma_{2}})_{m}|$ for every $m\geqs 0$.  
For every $k\geqs 0$, let $H_{k}$ denote the Hilbert function of the 
edge ideal of any $\Gamma\in \SE$ with $|Q|=k$.

Now, notice that if $\Gamma=([n],Q)\in \SE$ with $Q=\{i_{1}\!\cdot\!  
j_{1},\ldots,i_{k}\!\cdot\!  j_{k}\}$, then no monomial 
$x_{i_{t}}x_{j_{t}}$, for $t=1,\ldots,k$, is a zero divisor modulo the 
ideal $\langle 
x_{i_{1}}x_{j_{1}},\ldots,x_{i_{t-1}}x_{j_{t-1}}\rangle$. This implies, by 
\cite[\S 9.4, Cor.\ 5]{CLS} (or, rather, its proof) that
$$
H_{k+1}(m)=H_{k}(m)-H_{k}(m-2),\mbox{ for every } k\geqs 0,\ m\geqs 2,
$$
and hence
$$
H_{k+1}(m)=H_{0}(m)-\sum_{i=0}^{k}H_{i}(m-2),\mbox{ for every } k\geqs 0,\ m\geqs 
2;
$$
 we shall use this recursion to prove the expression in the statement by 
induction on $m$.

To begin with, we know  that 
$$
H_{0}(m)=|\MM(\underline{x})_{m}|=\binom{n+m}{n},\mbox{ for every } m\geqs 0,
$$
which clearly satisfies the expression in the statement (with $|Q|=0$).
Moreover,
$$
H_{k}(0)=1,\ H_{k}(1)=n+1\mbox{ for every } k\geqs 0,
$$
because, for 
every $\Gamma\in \SE$,
$$
C(I_{\Gamma})_{0}=\{1\},\ C(I_{\Gamma})_{1}=\{1,x_{1},\ldots,x_{n}\}.
$$
These values for $H_{k}(0)$ and $H_{k}(1)$ clearly satisfy the 
expression given in the statement. Now, as induction hypothesis, 
assume that
$$
H_{k}(m_{0})=\sum_{j=0}^{\lfloor m_{0}/2\rfloor} (-1)^j\binom{k}{j}
\binom{n+m_{0}-2j}{n} \mbox{ for every } k\geqs 0.
$$
Then, for every $k\geqs 0$,
$$
\begin{array}{rl}
H_{k+1}(m_{0}+2) &  =H_{0}(m_{0}+2)-\sum_{i=0}^{k}H_{i}(m_{0})\\
&  = \binom{n+m_{0}+2}{n}- \sum_{i=0}^{k}\sum_{j=0}^{\lfloor m_{0}/2\rfloor} 
(-1)^j\binom{i}{j}\binom{n+m_{0}-2j}{n}\\
&  =
\binom{n+m_{0}+2}{n}+\sum_{j=0}^{\lfloor m_{0}/2\rfloor} 
(-1)^{j+1}\binom{n+m_{0}-2j}{n}\Bigl(\sum_{i=0}^{k}\binom{i}{j}\Bigr)\\
&  =
\binom{n+m_{0}+2}{n}+\sum_{j=0}^{\lfloor m_{0}/2\rfloor} 
(-1)^{j+1}\binom{n+m_{0}-2j}{n}\binom{k+1}{j+1}\\
& =
(-1)^0\binom{k+1}{0}\binom{n+m_{0}+2}{n}+
\sum_{j=1}^{\lfloor m_{0}/2\rfloor+1} 
(-1)^{j}\binom{k+1}{j}\binom{n+m_{0}+2-2j}{n}\\
&  =
\sum_{j=0}^{\lfloor (m_{0}+2)/2\rfloor} 
(-1)^{j}\binom{k+1}{j}\binom{n+m_{0}+2-2j}{n},
\end{array}
$$
where the second equality uses the induction hypothesis and the fourth 
equality uses that $\sum_{i=0}^{k} \binom{i}{j}=\binom{k+1}{j+1}$.
\end{proof}

Thus, for instance, for every $\Gamma=([n],Q)\in \SE$,
$$
\begin{array}{rl}
H_{I_{\Gamma}}(2) &  =\binom{n+2}{2}-|Q|\\
H_{I_{\Gamma}}(3) &  =\binom{n+3}{3}-(n+1)|Q|\\
H_{I_{\Gamma}}(4)& =
\binom{n+4}{4}-\binom{n+2}{2}|Q|+\binom{|Q|}{2}\\
H_{I_{\Gamma}}(5)&  =
\binom{n+5}{5}-\binom{n+3}{3}|Q|+(n+1)\binom{|Q|}{2}\\
H_{I_{\Gamma}}(6)&  =
\binom{n+6}{6}-\binom{n+4}{4}|Q|+\binom{n+2}{2}\binom{|Q|}{2}-\binom{|Q|}{3}
\end{array}
$$

Unfortunately, we do not have a similar explicit expression for the 
Hilbert function of arbitrary contact structures, including unions of 
RNA secondary structures, and then we still use Lemma \ref{lema-N} to 
compute the Hilbert functions of the latter.

To close this paper, we shall provide explicit descriptions of 
$d_{5}$ and $d_{6}$ on the set $\SE$ just to grasp what they measure.  
Their proofs are simple, but long and technically involved, and we 
delay them until the Appendix at the end of this paper.

\begin{proposition}\label{d5-RNA}
For every $\Gamma_{1},\Gamma_{2}\in \SE$, 
$$
\begin{array}{l}
d_{5}(\Gamma_{1},\Gamma_{2})=|Q_{1}\Delta Q_{2}|\\
\qquad-\frac{1}{\binom{n+2}{2}}\Bigl(2(n-1)(|Q_{1}\Delta 
Q_{2}|-\Lambda_{\geq 2})+2\binom{|Q_{1}\cup 
Q_{2}|}{2}-\binom{|Q_{1}|}{2}-\binom{|Q_{2}|}{2}+2(\Lambda_{\geq 
3}+\Theta^{(4)})\Bigr)
\end{array}
$$
\end{proposition}

\begin{example}
Consider the RNA secondary structures of length 15
$$
\Gamma_{1}: (.).(.).(.).(.),\quad
\Gamma_{2}: ....(.(.).(.).),\quad
\Gamma_{3}: ..(.).(.).(.)..
$$
Then
$$
d_{3}(\Gamma_{1},\Gamma_{2})=d_{3}(\Gamma_{1},\Gamma_{3})=7
$$
and 
$$
d_{4}(\Gamma_{1},\Gamma_{2})=d_{4}(\Gamma_{1},\Gamma_{3})=\frac{25}{4},
$$
but
$$
d_{5}(\Gamma_{1},\Gamma_{2})= 5+\frac{71}{136},\quad
 d_{5}(\Gamma_{1},\Gamma_{3})=5+\frac{69}{136}.
$$
Thus, under $d_{5}$, $\Gamma_{3}$ is closer to $\Gamma_{1}$ than to 
$\Gamma_{2}$.
\end{example}

It is interesting to observe that, contrary to what happens with $d_{3}$ and $d_{4}$, 
the term $2\binom{|Q_{1}\cup 
Q_{2}|}{2}-\binom{|Q_{1}|}{2}-\binom{|Q_{2}|}{2}$ makes the value of 
$d_{5}(\Gamma_{1},\Gamma_{2})$ depend not only on the cardinal and 
structure of the set $Q_{1}\Delta 
Q_{2}$, but also on  $|Q_{1}\cap Q_{2}|$.  For instance, it is not difficult 
to check that if $\Gamma_{1},\Gamma_{2},\Gamma_{1}',\Gamma_{2}'\in 
\SE$ are such that $Q_{1}- Q_{2}=Q_{1}'- Q_{2}'$ and $Q_{2}- Q_{1}=Q_{2}'- Q_{1}'$, then
$$
d_{5}(\Gamma_{1},\Gamma_{2})<d_{5}(\Gamma'_{1},\Gamma'_{2})
\Longleftrightarrow |Q_{1}\cap Q_{2}|>|Q'_{1}\cap Q'_{2}|;
$$
i.e., the greater the set of contacts they share is, the closer they 
are. 

\begin{example}
Consider again the hairpin with an interior loop $\Gamma_{0}$ and its 
rearrangement $\Gamma_{1}$ given in Example \ref{ex-hairp}
$$
\Gamma_{0}: ((.((((...)))).....)),\quad \Gamma_{1}:
((..(((...)))(...).)).
$$
Let now $\Gamma'_{0}$ and $\Gamma'_{1}$ be the RNA secondary 
structures of the same length 21 obtained by removing from 
$\Gamma_{0}$ and $\Gamma_{1}$ their outer stacked pair of contacts:
$$
\Gamma'_{0}: ...((((...)))).......,\quad \Gamma'_{1}:
....(((...)))(...)....
$$
Since $Q_{1}\Delta Q_{2}=Q'_{1}\Delta Q'_{2}=\{4\!\cdot\!14,14\!\cdot 
\!18\}$, we have that
$$
d_{3}(\Gamma_{0},\Gamma_{1})=d_{3}(\Gamma'_{0},\Gamma'_{1})=2,\
d_{4}(\Gamma_{0},\Gamma_{1})=d_{4}(\Gamma'_{0},\Gamma'_{1})=\frac{21}{11}
$$
But 
it turns out that
$$
d_{5}(\Gamma_{0},\Gamma_{1})= 1+\frac{201}{253} ,\quad 
d_{5}(\Gamma'_{0},\Gamma'_{1})=
1+\frac{205}{253}.
$$
\end{example}

As far as $d_{6}$ goes, we have the following result.

\begin{proposition}\label{d6-RNA}
For every $\Gamma_{1},\Gamma_{2}\in \SE$, 
$$
\begin{array}{l}
d_{6}(\Gamma_{1},\Gamma_{2})=|Q_{1}\Delta Q_{2}| -\frac{1}{\binom{n+3}{3}}\Bigl((n+1)(2\binom{|Q_{1}\cup 
Q_{2}|}{2}- 
\binom{|Q_{1}|}{2}-\binom{|Q_{2}|}{2})\\
\qquad \qquad +2(\binom{n}{2}+3-|Q_{1}\cup 
Q_{2}|)(|Q_{1}\Delta Q_{2}|-\Lambda_{\geq 2})-2(n-1)\Lambda_{\geq 
3}+2\Lambda_{\geq 4}+2(n-3)\Theta^{(4)}
\Bigr)
\end{array}
$$
\end{proposition}

In a similar way, an explicit expression for $d_{m}$ on $\SE$ can be 
obtained for every $m\geqs 7$, yielding information about what these metrics 
measure: recall moreover that, for specific $\Gamma_{1},\Gamma_{2}\in \SE$, the 
value of $d_{m}(\Gamma_{1},\Gamma_{2})$ can be easily computed using a suitable 
computer algebra system.  Unfortunately, we have not been able to 
produce a closed expression for all these metrics.  Notice that, when 
finding an expression for $d_{m}$, the only new ingredient that is necessary to 
determine is the coefficient $SF_{m-1}(I_{\Gamma_{1}}\cup 
I_{\Gamma_{2}})$, which can be done for each $m$ by counting carefully 
how many square-free monomials of total degree $m-1$ belong to 
$I_{\Gamma_{1}}\cup I_{\Gamma_{2}}$ as we do in this paper for $m=4,5,6$.  It is 
in  this coefficient that new terms make their 
appearance in each $d_{m}$: when one balances the number of 
square-free monomials in $I_{\Gamma_{1}}\cup I_{\Gamma_{2}}$ of the 
form $x_{i_{1}}\cdots x_{i_{m-1}}$ such that $i_{1}\!\cdot \!  
i_{2},\ldots,i_{m-2}\!\cdot \!  i_{m-1}\in Q_{1}\cup Q_{2}$, the 
number $\Lambda^{(m-2)}$ makes its first 
appearance, 
and if $m-1$ is even, then to counterbalance the number of square-free 
monomials $x_{i_{1}}\cdots x_{i_{m-1}}$ in $I_{\Gamma_{1}\cup 
I_{\Gamma_{2}}}$ such that $\{i_{1},\ldots,i_{m-1}\}$ is a cyclic 
orbit, the number $\Omega^{(m-1)}$ must be used for the first time 
(cf.\ the proofs of Propositions \ref{d5-RNA} and \ref{d6-RNA} in the 
Appendix).

\section{Conclusion}

In the Discussion section of their paper \cite{RS96}, Reidys and 
Stadler, having pointed out that their group-based models and metrics 
cannot be used on arbitrary contact structures, ask ``What if contacts 
are not unique as in the case of proteins?''   Using edge ideals, we 
can represent arbitrary 
contact structures by means of monomial ideals of a polynomial ring, 
and we show that this representation generalizes the embedding of RNA secondary 
structures into the set of subgroups of $S_{n}$ proposed by Reidys and 
Stadler.  We have used then this representation to define a family of 
\emph{edge ideal metrics} on arbitrary contact structures, which can 
be easily computed using several freely available computer algebra 
systems, and we have studied their properties.

Edge ideals are not the unique possible monomial ideal representations 
of arbitrary contact structures.  For instance, we could associate to 
every contact structure $\Gamma=([n],Q)$ the \emph{clique ideal} 
$J_{\Gamma}$ of $\FF[x_{1},\ldots,x_{n}]$ generated by the set of 
monomials consisting of one square-free monomial $x_{i_{1}}\cdots 
x_{i_{k}}$ for each non-trivial \emph{clique} (complete subgraph) 
$\{i_{1},\ldots,i_{k}\}$, with $k\geqs 2$, of $\Gamma$.  Notice that 
if $\Gamma$ is an RNA secondary structure, then 
$J_{\Gamma}=I_{\Gamma}$, but for arbitrary contact structures they can 
be different. For instance, if $\Gamma=([5],\{1\!\cdot\!3,
3\!\cdot\!5,1\!\cdot\!5\})$, then
$$
I_{\Gamma}=\langle x_{1}x_{3},x_{1}x_{5},x_{3}x_{5}\qquad\rangle\mbox{ 
while }\qquad J_{\Gamma}=\langle x_{1}x_{3}x_{5}\rangle.
$$
We see that the clique ideal $J_{\Gamma}$ captures  information on 
the clusters of monomers in three-dimensional structures (for instance, base triplets and quartets in 
RNA structures) in a way different to $I_{\Gamma}$.  These ideals can be used to define new 
metrics on arbitrary contact structures of a fixed length similar to 
the edge ideal metrics introduced here.  We shall report on them in a 
subsequent paper.

Let us finally point out that another question of Reidys and Stadler's 
remains open for our models as well as, to our knowledge, for theirs: ``Is there any 
hope for extending or altering any of the above concepts in order to 
incorporate variable sizes of structures?'' 
\medskip

\noindent\textbf{Acknowledgments.} We acknowledge with thanks X. 
Bordoy, J. Elias, J. Mir\'o and G. Valiente for several discussions on the 
topic of this paper and for their comments on draft versions of it.

{\small\section*{Appendix: Proof of Propositions \ref{d5-RNA} and \ref{d6-RNA}}

To simplify the proofs, we establish first a lemma that we shall use 
several times and that generalizes the computation of 
$A(\Gamma_{1}\cup\Gamma_{2})$ carried on in the proof of Proposition 
\ref{prop-d4-RNA}.

For every $\Gamma_{1},\Gamma_{2}\in \SE$ and for every $k\geqs 2$, 
let
$$
M_{k}= \{\{i_{1}\!\cdot\! i_{2},i_{2}\!\cdot \! 
i_{3},\ldots,i_{k}\!\cdot\!i_{k+1}\}\subseteq\in Q_{1}\cup Q_{2}
\mid i_{1},\ldots,i_{k}\mbox{ pairwise different}\},
$$
and let $A_{k}$ be its cardinal.  Notice that $A_{2}$ is equal to the 
number of angles $A(\Gamma_{1}\cup I_{\Gamma_{2}})$ in $\Gamma_{1}\cup 
{\Gamma_{2}}$.  To simplify the notations, from now on we shall 
systematically write $A_{2}$ instead of $A(\Gamma_{1}\cup 
I_{\Gamma_{2}})$.
\medskip

\noindent\textbf{Lemma A}
For every $\Gamma_{1},\Gamma_{2}\in \SE$ and for every $k\geqs 2$, 
$$
A_{k}=|Q_{1}\Delta Q_{2}|-\sum_{m=4}^{k}m\Theta^{(m)}-\sum_{i=2}^{k} 
\Lambda_{\geq i}.
$$
\medskip

\begin{proof}
If $\{i_{1}\!\cdot\!  i_{2},,\ldots,i_{k}\!\cdot\!i_{k+1}\}\in M_{k}$, 
then the nodes $i_{1},i_{2},\ldots,i_{k+1}$ belong to the same orbit, 
whose length will be at least $k+1$.  Therefore, every cyclic orbit of 
length $m\leqs k$ contributes no element to $M_{k}$, while every cyclic 
orbit of length $m\geqs k+1$ adds $m$ new elements to it.  On 
the other hand, each linear orbit of length $m\leqs k$ contributes no 
element to $M_{k}$, while every linear orbit of length $m\geqs k+1$ 
adds $m-k$ new elements to this set.  

This shows that
$$
\begin{array}{rl}
A_{k}& =\sum_{m>k}m\Theta^{(m)}+\sum_{m>k}(m-k)\Lambda^{(m)}\\ 
&=\sum_{m>k}m\Theta^{(m)}+\sum_{m>k}(m-1)\Lambda^{(m)}
-(k-1)\sum_{m>k}\Lambda^{(m)}\\
& =|Q_{1}\Delta Q_{2}|-\sum_{m=4}^{k}m\Theta^{(m)}-\sum_{m=2}^{k} 
(m-1)\Lambda^{(m)}-(k-1)\sum_{m>k}\Lambda^{(m)}\\
& =|Q_{1}\Delta Q_{2}|-\sum_{m=4}^{k}m\Theta^{(m)}-\sum_{i=2}^{k} 
\Lambda_{\geq i}.
\end{array}
$$
\end{proof}

In particular, we obtain again that $A_{2}=|Q_{1}\Delta 
Q_{2}|-\Lambda_{\geq 2}$, as we already saw in the proof of 
Proposition \ref{prop-d4-RNA}.
\medskip

\noindent\emph{Proof of Proposition \ref{d5-RNA}.}
To simplify the notations, we shall denote each 
$SF_{i}(I_{\Gamma_{1}\cup\Gamma_{2}})$ simply by $SF_{i}$.  We shall 
use the expression
$$
d_{5}(\Gamma_{1},\Gamma_{2})=\frac{1}{\binom{n+2}{2}}\Bigl(H_{\Gamma_{1}}(4)+H_{\Gamma_{2}}(4)-
2H_{\Gamma_{1}\cup \Gamma_{2}}(4)\Bigr),
$$
where we already know that
$$
\begin{array}{l}
H_{{\Gamma_{i}}}(4)=
\binom{n+4}{4}-\binom{n+2}{2}|Q_{i}|+\binom{|Q_{i}|}{2},\quad i=1,2\\
H_{{\Gamma_{1}}\cup I_{\Gamma_{2}}}(4)=\binom{n+4}{n}-
(4SF_{1}+ 6SF_{2}+ 4SF_{3}+SF_{4})
\end{array}
$$
with
$$
SF_{1}=0,\ SF_{2}=|Q_{1}\cup Q_{2}|,\ 
SF_{3}=(n-2)|Q_{1}\cup Q_{2}|-A_{2};
$$
the value of $SF_{3}$ was obtained in the proof of Proposition 
\ref{prop-d4}: notice that $T(\Gamma_{1}\cup\Gamma_{2})=0$ and recall 
that $A(\Gamma_{1}\cup\Gamma_{2})=A_{2}$.  
It remains to compute $SF_{4}$:
\begin{itemize}
\item[(1)] For every $i\!\cdot\!  j\in Q_{1}\cup Q_{2}$, there are 
$\binom{n-2}{2}$ square free monomials $x_{i}x_{j}x_{k}x_{l}\in 
M(I_{\Gamma_{1}\cup\Gamma_{2}})$.  This makes 
$\binom{n-2}{2}|Q_{1}\cup Q_{2}|$ such monomials.

\item[(2)] Now, if $i\!\cdot \!j,k\!\cdot \!l\in Q_{1}\cup Q_{2}$ with 
$\{i,j\}\cap \{k,l\}=\emptyset$, then the monomial 
$x_{i}x_{j}x_{k}x_{l}$ is counted twice in (1).  Therefore, we must 
subtract $\binom{|Q_{1}\cup Q_{2}|}{2}-A_{2}$ to the value given in 
(1).

\item[(3)] If $i\!\cdot \!j,j\!\cdot \!k\in Q_{1}\cup Q_{2}$ form an 
angle in $\Gamma_{1}\cup\Gamma_{2}$, then for every $l\notin\{i,j,k\}$ 
the monomial $x_{i}x_{j}x_{k}x_{l}$ is counted twice in (1).  Thus, we 
must also subtract $(n-3)A_{2}$.

\item[(4)] If $i\!\cdot \!j,j\!\cdot \!k,k\!\cdot\!  l\in Q_{1}\cup 
Q_{2}$, with $i,j,k,l$ pairwise different, then the monomial 
$x_{i}x_{j}x_{k}x_{l}$ is counted three times in (1), then it is 
subtracted once in (2) and it is subtracted twice in (3).  Therefore, 
to retrieve these monomials, we must add $A_{3}$.

\item[(5)] Finally, if $i\!\cdot \!j,j\!\cdot \!k,k\!\cdot\!  l, 
l\!\cdot \!  i\in Q_{1}\cup Q_{2}$, so that if $\{i,j,k,l\}$ form a 
cyclic orbit of length $4$, then the monomial $x_{i}x_{j}x_{k}x_{l}$ 
is counted four times in (1), it is subtracted twice in (2), it is 
subtracted four more times in (3) and it is added four times in (4).  
To balance these operations, we must subtract $\Theta^{(4)}$.
\end{itemize}
In all, this shows that
$$
SF_{4}=\binom{n-2}{2}|Q_{1}\cup Q_{2}|-\binom{|Q_{1}\cup 
Q_{2}|}{2}-(n-4)A_{2}+ A_{3}-\Theta^{(4)}
$$
and hence
$$
H_{\Gamma_{1}\cup\Gamma_{2}}(4)=\binom{n+4}{4}-\binom{n+2}{2}|Q_{1}\cup 
Q_{2}|+\binom{|Q_{1}\cup Q_{2}|}{2}+nA_{2}- A_{3}+\Theta^{(4)}.
$$
A simple computation shows then that
$$
\begin{array}{rl}
d_{5}'(\Gamma_{1},\Gamma_{2})& =H_{\Gamma_{1}}(4)+H_{\Gamma_{2}}(4)-
2H_{\Gamma_{1}\cup \Gamma_{2}}(4)\\
&=2\binom{n+4}{4}-\binom{n+2}{2}(|Q_{1}|+|Q_{2}|)+\binom{|Q_{1}|}{2}+
\binom{|Q_{2}|}{2}\\
&\quad\qquad -2\Bigl(\binom{n+4}{4}-\binom{n+2}{2}|Q_{1}\cup 
Q_{2}|+\binom{|Q_{1}\cup Q_{2}|}{2}+nA_{2}-
A_{3}+\Theta^{(4)}\Bigr)\\
&=\binom{n+2}{2}|Q_{1}\Delta Q_{2}|-\Bigl(2\binom{|Q_{1}\cup 
Q_{2}|}{2}- \binom{|Q_{1}|}{2}-\binom{|Q_{2}|}{2}+2nA_{2}- 
2A_{3}+2\Theta^{(4)}\Bigr)
\end{array}
$$
Now, we know from Lemma A that
$$
A_{2}=|Q_{1}\Delta Q_{2}|-\Lambda_{\geq 2},\quad
A_{3}=|Q_{1}\Delta Q_{2}|-\Lambda_{\geq 2}-\Lambda_{\geq 3}.
$$
Replacing them in the expression obtained above for 
$d_{5}'(\Gamma_{1},\Gamma_{2})$, and dividing the resulting expression 
by $\binom{n+2}{2}$, we finally obtain
$$
\begin{array}{l}
d_{5}(\Gamma_{1},\Gamma_{2})=|Q_{1}\Delta Q_{2}|\\
\qquad-\frac{1}{\binom{n+2}{2}}\Bigl(2\binom{|Q_{1}\cup 
Q_{2}|}{2}-\binom{|Q_{1}|}{2}-\binom{|Q_{2}|}{2}+2(n-1)(|Q_{1}\Delta 
Q_{2}|-\Lambda_{\geq 2})+2(\Lambda_{\geq 3}+\Theta^{(4)})\Bigr),
\end{array}
$$
as we wanted to prove.\hfill \qed
\medskip

\noindent\emph{Proof of Proposition \ref{d6-RNA}.} To simplify the 
notations, we shall denote again 
$SF_{i}(I_{\Gamma_{1}\cup\Gamma_{2}})$ simply by $SF_{i}$.  In the 
expression
$$
d_{6}(\Gamma_{1},\Gamma_{2})=\frac{1}{\binom{n+3}{3}}\Bigl(H_{\Gamma_{1}}(5)+H_{\Gamma_{2}}(5)-
2H_{\Gamma_{1}\cup \Gamma_{2}}(5)\Bigr),
$$
 we already know that
$$
\begin{array}{l}
H_{{\Gamma_{i}}}(5)=
\binom{n+5}{5}-\binom{n+3}{3}|Q_{i}|+(n+1)\binom{|Q_{i}|}{2},\quad 
i=1,2\\
H_{{\Gamma_{1}}\cup I_{\Gamma_{2}}}(5)=\binom{n+5}{n}-
(5SF_{1}+ 10SF_{2}+ 10SF_{3}+5SF_{4}+SF_{5})
\end{array}
$$
with
$$
\begin{array}{c}
SF_{1}=0,\ SF_{2}=|Q_{1}\cup Q_{2}|,\
SF_{3}=(n-2)|Q_{1}\cup Q_{2}|-A_{2}\\
SF_{4}=\binom{n-2}{2}|Q_{1}\cup Q_{2}|-\binom{|Q_{1}\cup 
Q_{2}|}{2}-(n-4)A_{2}+ A_{3}-\Theta^{(4)}
\end{array}
$$
Let us compute now $SF_{5}$:
\begin{itemize}
\item[(1)] For every $i\!\cdot\!  j\in Q_{1}\cup Q_{2}$, there are 
$\binom{n-2}{3}$ square free monomials $x_{i}x_{j}x_{k}x_{l}x_{m}\in 
M(I_{\Gamma_{1}\cup\Gamma_{2}})$.  This makes 
$\binom{n-2}{3}|Q_{1}\cup Q_{2}|$ such monomials.

\item[(2)] Now, if $i\!\cdot \!j,k\!\cdot \!l\in Q_{1}\cup Q_{2}$ with 
$\{i,j\}\cap \{k,l\}=\emptyset$, then for every 
$m\notin\{i,j,k,l\}$ the monomial $x_{i}x_{j}x_{k}x_{l}x_{m}$ is 
counted twice in (1).  Therefore,  we must subtract 
$(n-4)\Bigl(\binom{|Q_{1}\cup Q_{2}|}{2}-A_{2}\Bigr)$ to (1).

\item[(3)] If $i\!\cdot \!j,j\!\cdot \!k\in Q_{1}\cup Q_{2}$, then for 
every $l,m\notin\{i,j,k\}$ the monomial $x_{i}x_{j}x_{k}x_{l}x_{m}$ is 
counted twice in (1).  Therefore, we must also subtract 
$\binom{n-3}{2}A_{2}$ to (1).

\item[(4)] If $i\!\cdot \!j,j\!\cdot \!k,l\!\cdot\!  m\in Q_{1}\cup 
Q_{2}$ with $\{i,j,k\}\cap\{l,m\}=\emptyset$, the monomial 
$x_{i}x_{j}x_{k}x_{l}x_{m}$ is counted 3 times in (1), then it is 
subtracted twice in (2) and it is subtracted once again in (3).  
Therefore, to retrieve these monomials we must add
$$
\Bigl|\Bigl\{\{i\!\cdot \!j,j\!\cdot \!k,l\!\cdot\! m\}\mid 
i\!\cdot \!j,j\!\cdot \!k,l\!\cdot\! m\in Q_{1}\cup Q_{2}, 
\{i,j,k\}\cap\{l,m\}=\emptyset\Bigr\}\Bigr|;
$$ 
let us call for the moment $X$ this number.

\item[(5)] If $i\!\cdot \!j,j\!\cdot \!k,k\!\cdot\!  l\in Q_{1}\cup 
Q_{2}$, for every $m\notin\{i,j,k,l\}$ the monomial 
$x_{i}x_{j}x_{k}x_{l}x_{m}$ is counted 3 times in (1), then it is 
subtracted once in (2) and it is subtracted twice in (3).  Therefore, 
to retrieve these monomials we must also add $(n-4)A_{3}$ monomials.

Notice now that 
$$
X+A_{3}=\Bigl|\Bigl\{\{i\!\cdot \!j,j\!\cdot \!k,l\!\cdot\! m\}\mid 
i\!\cdot \!j,j\!\cdot \!k,l\!\cdot\! m\in Q_{1}\cup 
Q_{2}\Bigr\}\Bigr|=(|Q_{1}\cup Q_{2}|-2)A_{2}.
$$
Therefore, (4) and (5) add jointly
$$
(|Q_{1}\cup Q_{2}|-2)A_{2}+(n-5)A_{3}
$$
monomials.

\item[(6)]  If $i\!\cdot \!j,j\!\cdot \!k,k\!\cdot\!  l, l\!\cdot \!  
i\in Q_{1}\cup Q_{2}$, i.e., if $\{i,j,k,l\}$ form a cyclic orbit of 
length $4$, then for every $m\notin\{i,j,k,l\}$
 the monomial $x_{i}x_{j}x_{k}x_{l}x_{m}$ is counted four 
times in (1), it is subtracted twice in (2), it is subtracted 
four more times in (3) and it is added four times in (5). Therefore, we 
must subtract $(n-4)\Theta^{(4)}$.

\item[(7)] Finally, if $i\!\cdot \!j,j\!\cdot \!k,k\!\cdot\!  l, 
l\!\cdot \!  m\in Q_{1}\cup Q_{2}$, then the monomial 
$x_{i}x_{j}x_{k}x_{l}x_{m}$ is counted four times in (1), it is 
subtracted three times in (2), it is subtracted three more times in 
(3) and it is added twice in (4) and twice in (5).  Therefore, we must 
subtract $A_{4}$.
\end{itemize}

In all, this shows that
$$
\begin{array}{l}
SF_{5} =\binom{n-2}{3}|Q_{1}\cup 
Q_{2}|-(n-4)(\binom{|Q_{1}\cup Q_{2}|}{2}-A_{2})-\binom{n-3}{2}A_{2}\\
\qquad\qquad\qquad +(|Q_{1}\cup 
Q_{2}|-2)A_{2}+(n-5)A_{3}-(n-4)\Theta^{(4)}-A_{4}
\end{array}
$$
and hence
$$
\begin{array}{l}
H_{\Gamma_{1}\cup\Gamma_{2}}(5)=\binom{n+5}{5}-\binom{n+3}{3}|Q_{1}\cup 
Q_{2}|+(n+1)\binom{|Q_{1}\cup Q_{2}|}{2}+(\binom{n+1}{2}-|Q_{1}\cup 
Q_{2}|+2)A_{2}\\
\qquad\qquad\qquad -nA_{3}+(n+1)\Theta^{(4)}+A_{4}.
\end{array}
$$
Then
$$
\begin{array}{l}
d_{6}'(\Gamma_{1},\Gamma_{2})=H_{\Gamma_{1}}(5)+H_{\Gamma_{2}}(5)-
2H_{\Gamma_{1}\cup \Gamma_{2}}(5)\\
\quad=2\binom{n+5}{5}-\binom{n+3}{3}(|Q_{1}|+|Q_{2}|)+
(n+1)(\binom{|Q_{1}|}{2}+\binom{|Q_{2}|}{2})\\
\qquad -2\Bigl(
\binom{n+5}{5}-\binom{n+3}{3}|Q_{1}\cup 
Q_{2}|+(n+1)\binom{|Q_{1}\cup Q_{2}|}{2}\\
\qquad\qquad +(\binom{n+1}{2}-|Q_{1}\cup 
Q_{2}|+2)A_{2}-nA_{3}+(n+1)\Theta^{(4)}+A_{4}\Bigr)\\
\quad=\binom{n+3}{3}|Q_{1}\Delta Q_{2}|-(n+1)\Bigl(2\binom{|Q_{1}\cup 
Q_{2}|}{2}-
\binom{|Q_{1}|}{2}-\binom{|Q_{2}|}{2}\Bigr)-2\Bigl(\binom{n+1}{2}+2-|Q_{1}\cup 
Q_{2}|\Bigr)A_{2}\\
\qquad\qquad
+2nA_{3}-2(n+1)\Theta^{(4)}-2A_{4}.
\end{array}
$$
Since we already know, by Lemma A,
that
$$
\begin{array}{c}
A_{2}=|Q_{1}\Delta Q_{2}|-\Lambda_{\geq 2},\quad A_{3}= |Q_{1}\Delta 
Q_{2}|-\Lambda_{\geq 2}- \Lambda_{\geq 3}\\
A_{4}= |Q_{1}\Delta 
Q_{2}|-\Lambda_{\geq 2}- \Lambda_{\geq 3}-\Lambda_{\geq 
4}-4\Theta^{(4)}
\end{array}
$$
when we replace these values in the expression obtained above for 
$d_{6}'(\Gamma_{1},\Gamma_{2})$ and we divide the resulting expression 
by $\binom{n+3}{3}$, we finally obtain
$$
\begin{array}{l}
d_{6}(\Gamma_{1},\Gamma_{2})=|Q_{1}\Delta Q_{2}| -
\frac{1}{\binom{n+3}{3}}\Bigl((n+1)(2\binom{|Q_{1}\cup 
Q_{2}|}{2}- 
\binom{|Q_{1}|}{2}-\binom{|Q_{2}|}{2})\\
\qquad \qquad +2(\binom{n}{2}+3-|Q_{1}\cup 
Q_{2}|)(|Q_{1}\Delta Q_{2}|-\Lambda_{\geq 2})-2(n-1)\Lambda_{\geq 
3}+2\Lambda_{\geq 4}+2(n-3)\Theta^{(4)}
\Bigr)
\end{array}
$$
\ \hfill \qed
}
\end{document}